\documentclass[acmsmall]{acmart}
\makeatletter
\providecommand\ACM@cc@type{by}
\makeatother
\AtBeginDocument{%
  }
\usepackage{subcaption}
\usepackage{array}
\acmJournal{TKDD}
\acmVolume{0}
\acmNumber{0}
\acmArticle{0}
\acmYear{2026}
\acmMonth{8}
\setcopyright{none}
\copyrightyear{2026}
\acmYear{2026}
\acmDOI{XXXXXXX.XXXXXXX}

\begin{document}

\title{Scalable dynamic community detection on temporal graphs using graph neural networks}


\author{Peijie Zhong}
\affiliation{%
  \institution{Queen Mary University of London}
  \city{London}
  \country{United Kingdom}}
\email{p.zhong@qmul.ac.uk}

\author{Ra\'ul Mondrag\'on}
\affiliation{%
  \institution{Queen Mary University of London}
  \city{London}
  \country{United Kingdom}
}\email{r.j.mondragon@qmul.ac.uk}

\author{Richard Clegg}
\affiliation{%
  \institution{Queen Mary University of London}
  \city{London}
  \country{United Kingdom}}
\email{r.clegg@qmul.ac.uk}
\renewcommand{\shortauthors}{Zhong et al.}

\begin{abstract} 
Dynamic community detection on temporal graphs seeks to identify evolving community structures while allowing node memberships to change over time. In this work, we formulate dynamic community detection over observed node–time instances, where each node-time instance in the temporal interaction stream is assigned a cluster label. We propose a diffusion-guided contrastive learning framework that uses a local temporal diffusion affinity matrix to construct positive and negative node–time pairs and organise the learned representations according to their temporal structural relationships. We then apply a clustering algorithm to the resulting embedding space to detect dynamic communities. Experiments on synthetic temporal networks show that the proposed method outperforms static community detection baselines and achieves competitive or better performance than existing dynamic community detection methods in terms of AMI and ARI, while maintaining good scalability. We further apply the method to a large-scale OpenAlex computer science collaboration network from 2016 to 2025, revealing persistent and evolving collaboration communities in real scientific data. These results suggest that time-node-level representation learning provides an effective framework for scalable dynamic community detection on temporal graphs.
\end{abstract}
\acmDOI{}
\acmISBN{}
\acmPrice{}



\maketitle

\section{Introduction}

\begin{figure}
    \centering
    \includegraphics[width=0.9\linewidth]{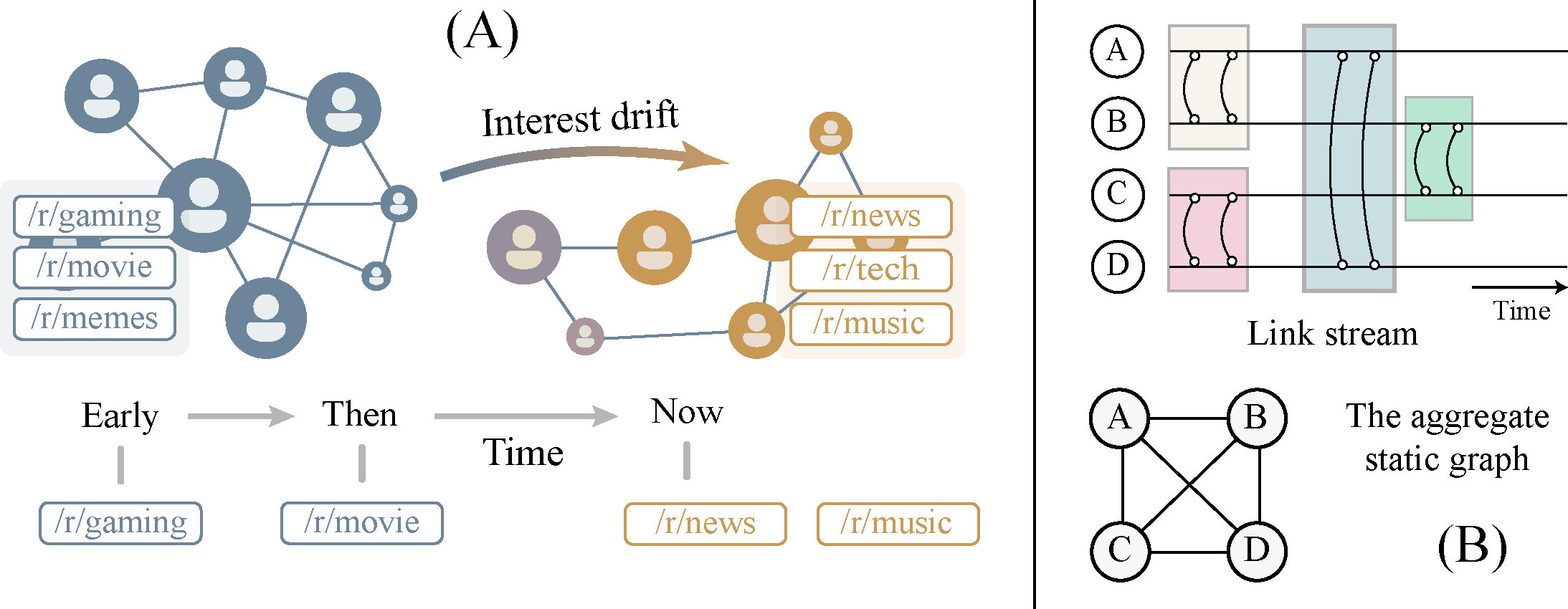}
    \caption{(A) demonstrates the phenomenon of user interest drift on a social platform. Users’ interests may change over time, leading to evolving community membership. (B) aggregating a link stream into a static graph may obscure temporal changes in community structure.}
    \label{fig:introduction}
\end{figure}

Many real-world networks are inherently temporal, with interactions evolving continuously over time rather than remaining fixed. Social networks, communication systems, collaboration networks, and transaction networks are often represented as temporal graphs or link streams, where edges appear, disappear, and recur at different timestamps. In such networks, the structural role of a node may change over time. For example, as illustrated in Fig.~\ref{fig:introduction}~(A), a user on a social platform may initially participate in one topic group and later shift attention to other topics. This type of temporal variation suggests that assigning one fixed community label to each node over the entire observation period is often insufficient for analysing evolving temporal networks. As illustrated in Fig.~\ref{fig:introduction}~(B), aggregating a temporal graph into a static graph may obscure temporal changes in community structure. This motivates dynamic community detection, where the goal is to recover not only groups of related nodes, but also how these groups and node memberships evolve over time. A half-way approach is given by snapshot based methods~\cite{mucha2010community, bazzi2016community} which, instead of treating the network as a truly temporal network, work on an ordered series of static networks defined by time windows. Recently, Longitudinal Agglomerative Greedy Optimization (LAGO)~\cite{brabant2025discovering} was the first genuinely dynamic community detection method. It introduce the concept of a node-time set which defines the community for a node at each point in time where it interacts with another node. However, the scaling behaviour of LAGO is poor and it struggles with real-world network sizes.

This paper is the first to demonstrate a method that can provide truly dynamic community detection in large-scale temporal networks. Unlike LAGO~\cite{brabant2025discovering}, which directly optimises community assignments over the node-time set, our method uses a learning-based framework to make dynamic community detection more scalable on large temporal networks. We validate the method by using synthetic networks with a known ground-truth and demonstrate that we can recover the ground truth partitions. We compare our method with LAGO~\cite{brabant2025discovering}, static methods~\cite{nguyen2018continuous, blondel2008fast, Liu2025DeepDatasets} and snapshot-based dynamic methods~\cite{mucha2010community, bazzi2016community} demonstrating that we better match the ground truth. We measure the quality of detected dynamic community structure using adjusted mutual information (AMI)~\cite{vinh2009information, danon2005comparing} and adjusted Rand index (ARI)~\cite{rand1971objective, hubert1985comparing}, two established metrics that can compare community assignments that change in time.
We also demonstrate that we scale better in terms of run-time as network size increases (the only algorithm with comparable run-time scaling has much worse performance against the ground truth). We demonstrate the utility of our algorithm on real-world datasets by running it on a large-scale network of computer science collaborations known as OpenAlex~\cite{priem2022openalex} and show that we can interpret how communities evolve and split as the network changes.

In order to provide truly dynamic community analysis, our method needs to keep the fine temporal resolution of link streams while simultaneously remaining scalable on large networks. To this end, we introduce a graph representation learning framework for dynamic community detection. Instead of assigning one label to each node, the proposed framework assigns a community label to each node-time instance, by which we mean an observed occurrence of a node at a particular timestamp in the link stream. In this way, the same node can receive different community labels when it appears in different temporal interaction contexts. This allows the same node to belong to different communities at different timestamps. To improve scalability, the model learns low-dimensional node-time representations with a temporal graph encoder and optimises the objective in mini-batches, rather than directly searching over all node-time community assignments. To learn representations tailored to dynamic community detection, we introduce a diffusion-guided contrastive learning framework. At each timestamp, a local temporal diffusion affinity matrix captures the structural relationships between node–time instances and is used to construct positive and negative training pairs. The resulting contrastive objective brings structurally related instances closer while pushing unrelated ones apart, thereby embedding local temporal structure into the representation space. We then apply a clustering algorithm to the embedding space produced by the trained model to detect dynamic community structures.
\section{Related work}
\begin{table}[t]
    \centering
    \begin{tabular}{l l l c}
    \toprule
    Name & Classification & Strategy & Complexity\\
    \midrule
    Louvain~\cite{blondel2008fast} & Static & Modularity maximisation & $O(N\log(N))$\\
    CTDNE~\cite{nguyen2018continuous}+K-Means & Static & Random walk & $O(\rho lNd)$\\
    TGC~\cite{Liu2025DeepDatasets} & Static & Graph neural network & $O(rE)$\\
    MAGI~\cite{liu2024revisiting} & Static & Graph neural network & $O(Nrd^2)$\\
    GenLouvain~\cite{mucha2010community, bazzi2016community} & Window-based & Modularity maximisation & $O(NL)$ \\
    LAGO~\cite{brabant2025discovering} & Dynamic & Modularity maximisation & $O(EN^2)$\\
        \bottomrule
    \end{tabular}
    \caption{Comparison of representative static and dynamic clustering/community detection methods used in this study. Static clustering methods assign one fixed cluster label to each node over the whole observation period, whereas dynamic clustering methods allow node memberships to vary over time. The table summarises the main modelling strategy and asymptotic complexity of each method. Here, $N$ denotes the number of nodes, $E$ denotes the number of temporal interactions or edges, $d$ denotes the embedding dimension, $l$ denotes the random-walk length. $\rho$ denotes the number of random walks sampled per node, $r$ denotes the number of training epochs for graph neural networks, and $L$ denotes the number of layers after the temporal network is discretised into a multilayer representation.}
    \label{tab:Algorithms}
\end{table}

This section reviews prior work related to dynamic community detection on temporal graphs. We first clarify the connection between node clustering and community detection, since the proposed method uses a learning-based clustering framework to recover evolving community structure. We then review six representative baseline algorithms, summarised in Table~\ref{tab:Algorithms}. These methods fall into three classes according to how node memberships are allowed to change over time: static methods (Louvain, CTDNE+K-Means, TGC, MAGI) assign each node a single community label for the entire observation period; the window-based method (GenLouvain) discretises the temporal network into a sequence of layers and detects communities jointly across this multilayer representation; and the dynamic method (LAGO) allows node memberships to vary over time. For each method, Table~\ref{tab:Algorithms} also reports the underlying strategy: modularity maximisation, random-walk-based embedding, and graph neural networks, together with the asymptotic computational complexity, allowing the scalability of different approaches to be compared directly. In the rest of this section, we discuss these static and dynamic community detection methods in turn. In this part, we first review clustering on graphs based on nodes' representations, which provide the neural encoding mechanisms for time-evolving interaction data but are often optimised for predictive tasks rather than community recovery. Secondly, we discuss static and dynamic community detection methods, including snapshot-based and link-stream approaches, and highlight their limitations in modelling fine-grained temporal community evolution at scale.

\subsection{Node clustering and community detection in graphs}

Graph clustering and community detection are closely related problems in graph analysis. Graph clustering is often described from a machine-learning perspective, where nodes are partitioned according to similarity in a feature or embedding space. Such features may be external node attributes, or representations learned from the graph structure itself~\cite{cai2018comprehensive}. Community detection is usually described from a network-science perspective, where the goal is to identify groups of nodes with coherent connectivity patterns, such as dense internal connections relative to external connections. These two perspectives are not mutually exclusive. Spectral clustering~\cite{ng2001spectral, von2007tutorial} provides a classical example: it constructs a graph Laplacian from the network topology, uses its eigenvectors as low-dimensional node representations, and then applies a clustering algorithm such as K-Means in the resulting representation space. The resulting partitions can also be interpreted as communities in the original graph. More generally, when graph representations preserve the structural relationships that define communities, clustering over those representations can serve as a learning-based route to community detection. This connection provides the conceptual basis for using representation learning and clustering techniques to recover community structure, which is the perspective adopted in this paper for temporal graphs.

The work here might be argued to fit either into clustering or into community detection. Our explicit goal at the outset of the work was within the context of community detection: we evaluate our work using a dynamic variant of the community measure modularity and test it on synthetic networks that generate communities. We demonstrate its utility in finding communities within a network. However, essential parts of the mechanism within the work rely on clustering within a state-space, so the method also partially falls into the category of clustering algorithms. 

\subsection{Representation-based graph clustering}
Deep graph clustering aims to divide nodes into clusters by leveraging deep neural networks to learn structure- and attribute-aware node representations~\cite{liu2026survey}. Early methods such as graph autoencoder-based models learn node embeddings by reconstructing graph structure or node attributes, and then apply K-means or spectral clustering~\cite{wang2017mgae}. However, reconstruction objectives mainly encourage embeddings to preserve the observed graph structure or node attributes, rather than explicitly enforcing compact intra-cluster representations and separable inter-cluster boundaries. Later methods introduce clustering-oriented objectives, such as soft assignment refinement~\cite{wang2019attributed, bo2020structural}, to make node embeddings more compact within clusters and more separable across clusters. Recent contrastive methods further improve discriminability by constructing positive and negative pairs from graph augmentations, neighbourhood relations, and then pulling likely same-cluster nodes together while pushing different-cluster nodes apart~\cite{hassani2020contrastive, liu2024revisiting, liu2022deep, yang2023cluster}. MAGI~\cite{liu2024revisiting} reveals the strong connections between modularity maximisation and graph contrastive learning, where positive and negative examples are naturally defined by modularity, and uses this insight to design a community-aware graph clustering method that leverages modularity maximisation as a contrastive learning task to detect the underlying communities in static graphs. Despite the effectiveness, these methods are mainly designed for static graphs, where each node has a single representation and a single cluster assignment. They therefore cannot directly handle temporal graphs, in which interactions arrive continuously and the same node may change its community membership over time. 

Temporal graph representation learning aims to learn node representations from time-evolving interaction data. Depending on how time is represented, existing methods are commonly divided into discrete-time and continuous-time models. Discrete-time methods represent a temporal graph as a sequence of snapshots and combine static graph encoders with temporal modules to capture dependencies across layers. Representative examples include EvolveGCN~\cite{Pareja2019EvolveGCN} and DySAT~\cite{sankar2018dynamic}, which update node representations across graph snapshots using recurrent or attention-based mechanisms. Although these methods can model temporal evolution across snapshots, the aggregation of interactions into discrete layers inevitably loses fine-grained temporal information, such as event ordering, exact timestamps, and short-term interaction patterns.

Continuous-time temporal graph methods instead operate directly on timestamped interactions. Temporal random-walk methods such as Continuous-Time Dynamic Network Embeddings (CTDNE)~\cite{nguyen2018continuous} and Causal Anonymous Walks (CAW)~\cite{makarov2021temporal} preserve temporal ordering by sampling time-respecting walks from interaction sequences. Neural temporal graph models such as Temporal Graph Attention Network (TGAT)~\cite{da2020inductive} and TGN~\cite{emanuele2020temporal} learn time-dependent node representations through temporal neighbourhood aggregation, attention mechanisms, memory modules, or time encodings. Other methods, such as HTNE~\cite{zuo2018embedding} and MNCI~\cite{liumeng2021inductive}, model temporal dependency through point-process-based mechanisms. These methods provide powerful tools for encoding temporal interaction patterns, and they are closely related to our encoder design. However, most temporal graph representation learning methods are designed for predictive tasks, especially link prediction. Their training objectives encourage embeddings to preserve information useful for predicting future interactions, but they do not explicitly optimise the representation space for recovering evolving clustering structure. As a result, applying clustering directly to such embeddings may not produce meaningful dynamic clusters. Temporal Graph Clustering (TGC)~\cite{Liu2025DeepDatasets} is closer to our setting because it combines temporal graph representation learning with clustering. Nevertheless, it still assumes that each node has a single static cluster label over the whole observation period. This makes it unsuitable for settings where the same node may change its cluster label over time. In contrast, our method assigns labels to observed node occurrences in the temporal interaction stream and introduces local temporal structural alignment so that the learned node-time representations are guided by cluster-relevant temporal-topological information.
\subsection{Dynamic community detection}
Community detection in static graphs can be approached from several distinct principles. One class of methods is based on optimising partition quality functions, such as modularity~\cite{newman2006modularity}, which compares the observed within-community connectivity with that induced by a null model. Another important class includes spectral methods derived from graph Laplacians~\cite{von2007tutorial}. Probabilistic approaches, notably stochastic block models~\cite{holland1983stochastic, xu2014dynamic}, represent communities through latent generative structure, while random-walk-based methods~\cite{pons2005computing, smiljanic2026community} interpret communities as regions that retain diffusion flow over time. These formulations differ in their underlying assumptions, but each provides a structural criterion for partitioning nodes into coherent groups. Extending community detection to temporal graphs is more challenging because communities may evolve, split, merge, appear, or disappear over time. 

A common strategy is to discretise a temporal network into a sequence of snapshots. The most direct approach applies a static community-detection algorithm to each snapshot and subsequently matches or smooths the resulting partitions across time. Rather than recomputing each partition independently, paper~\cite{aynaud2010static} initialises Louvain at snapshot $t$ with the partition obtained at $t-1$, thereby improving temporal stability while retaining the local-moving procedure of static Louvain. Paper~\cite{chong2013incremental} similarly reuses the preceding partition, but formulate the update as a batch operation capable of accommodating multiple simultaneous changes. Paper~\cite{meng2016novel} also formulates the problem at the snapshot level, jointly optimising a weighted combination of the modularity at $t+1$ and at 
$t$ under a first-order Markov assumption. Paper~\cite{cordeiro2016dynamic} preserves unaffected communities and reopens only communities incident to added or removed vertices and edges. Delta-Screening~\cite{zarayeneh2021delta} identifies a candidate set of vertices whose modularity-maximising assignments may have changed and confines subsequent optimisation to this screened region. Dynamic Frontier~\cite{sahu2024df} starts from endpoints directly affected by a batch update and expands the active region only when a vertex changes community, thereby capturing cascading effects without processing the entire graph. 

Another snapshot-based formulation models the temporal network as an ordered multilayer graph and optimises objectives such as multilayer modularity~\cite{mucha2010community}. In this setting, each snapshot corresponds to one layer and inter-layer couplings encourage temporal coherence, often by penalising excessive community switching across adjacent layers~\cite{bazzi2016community, pamfil2019relating}. Similar ideas also appear in multilayer random-walk formulations~\cite{aslak2018constrained}, which extend the random walk to allow walker to jump across layers. These approaches jointly capture within-layer structure and across-layer persistence, but they still rely on prior discretisation into snapshots and often assume relatively well-aligned node identities across layers. When the node set changes substantially over time, additional treatment is required, which may complicate both modelling and optimisation. 

These methods make snapshot-based community detection computationally practical, but the resulting communities remain conditional on the temporal discretisation: the window width determines which interactions are aggregated, while the window boundaries determine which changes are presented to the algorithm as a single batch.

To avoid the aforementioned limitations of snapshot-based model, recent work has turned to the link-stream perspective, where the temporal network is represented directly as a sequence of timestamped interactions. In this setting, community structure is defined over continuous time rather than over pre-aggregated windows, making it possible to preserve fine-grained temporal information and capture short-lived interaction patterns. Longitudinal modularity~\cite{brabant2025longitudinal} is an example of such a formulation, extending modularity to link streams by evaluating community assignments directly on temporal interaction data. Based on this metric, link-stream community detection algorithms such as LAGO~\cite{brabant2025discovering} search for partitions over time-nodes, and iteratively improve the objective through local moves and temporal restructuring in a Louvain-style folding. The main strength of this line of work is that it avoids human-chosen windows and is more faithful to the temporal resolution of the data. However, this benefit comes with a substantial increase in the size of the optimisation space, since the temporal dimension effectively lifts the problem from node partitions to node-time partitions. Consequently, although link-stream methods provide a principled framework for fine-grained dynamic community detection, their computational cost can become prohibitive on large-scale temporal networks or high-frequency interaction streams. 

Existing studies provide useful components for modelling temporal networks, but they do not fully address the setting considered in this paper. Temporal graph representation learning offers powerful encoders for temporal interactions, yet its objectives are often designed for prediction rather than community recovery. Snapshot-based dynamic community detection captures temporal evolution more directly, but it relies on predefined aggregation windows. Link-stream modularity optimisation avoids such discretisation, but its search space grows substantially when community assignments are defined over node-time instances. The proposed method combines these perspectives by learning low-dimensional node-time representations, guiding them with local temporal random-walk structure, and optimising them in mini-batches for scalable dynamic community detection. 
\section{Method}
In this section, we present the proposed dynamic clustering framework. The related code and experiments are available on GitHub~\footnote{\url{https://github.com/Peijie-Zhong/Dycomm-detection}}. We first define continuous-time temporal graphs as link streams and formulate the task as clustering observed node-time instances. We then describe the model architecture, which integrates temporal representation learning with diffusion-based structural supervision to learn community-aware node-time embeddings.

\subsection{Problem definition}

Here we give the definition of temporal graph and dynamic community on temporal graphs. 
\paragraph{Definition 1. Temporal graphs:} Let $G=(V,E,T)$ denote a temporal interaction network represented as a link stream, where $V$ is the set of nodes and $E=\{(u_i,v_i,t_i)\}_{i=1}^{|E|}$ is the set of timestamped interactions. Each interaction $(u_i,v_i,t_i)$ indicates that nodes $u_i, v_i \in V$ interact at time $t_i \in \mathbb{R}$. The set of timestamps is defined as $T=\{t_i \mid \exists (u_i,v_i,t_i)\in E\}$. We further use
\begin{equation*}
    \mathcal{V}=\{(u,t)\mid \exists v\in V \text{ such that } (u,v,t)\in E\}
\end{equation*}
to denote the set of node-time instances, namely, the instances for which node $u$ participates in at least one interaction at time $t$. Unlike static graphs, the topology of a temporal graph is implicitly defined by the time-ordered interactions rather than by a single fixed adjacency matrix.
\paragraph{Definition 2. Dynamic communities in temporal graphs:} A dynamic community is an assignment of community labels to node-time instances. This allows the community structure to change over time. Specifically, instead of assigning a fixed label to each node $u\in V$, we assign a label to each observed pair $(u,t)$, where $u$ is a node and $t$ is a time point at which the node appears in the link stream as part of some timestamped interaction, $(u_i,v_i,t_i)$. Formally, the goal is to learn a mapping $c:\mathcal{V}\rightarrow \{1,2,\dots,K\}$, where $\mathcal{V}$ denotes the set of observed node-time instances and $K$ is the number of communities. This formulation allows a node to take different community labels at different times, thus characterising the dynamic evolution of community structure.

Several criteria could be used to evaluate the quality of such a mapping. For synthetic datasets, the planted community assignments provide a ground truth against which the inferred communities can be compared directly. In contrast, real-world temporal networks typically lack ground-truth community labels, making evaluation considerably more challenging. One possible approach is to assess goodness-of-fit with respect to a generative model; however, unlike the static setting, there is currently no widely adopted generative model for community structure in continuous-time networks~\cite{yuan2025temporal, brabant2025longitudinal, agdur2025approximating}. Therefore, in this work, we evaluate community assignments on real-world datasets using the longitudinal modularity~\cite{brabant2025longitudinal}, which quantifies the extent to which interactions are concentrated within communities over time in a continuous-time temporal graph.

\subsection{Model structure}

\begin{figure}
    \centering
    \includegraphics[width=\linewidth]{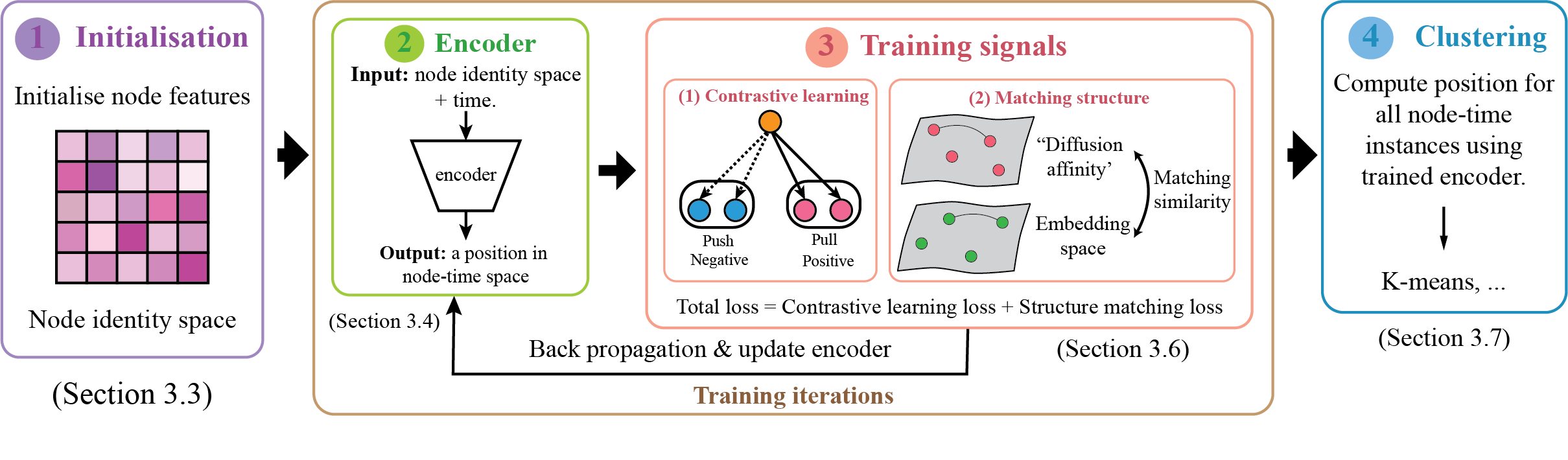}
    \caption{Overview of the proposed dynamic clustering framework. Node identity features are first initialised and mapped into a node-time embedding space by a temporal graph encoder. The encoder is trained using contrastive learning and diffusion-based structural matching. The resulting node-time embeddings are then partitioned to obtain dynamic community assignments.}
    \label{fig:structure}
\end{figure}

The overall structure of the model is shown in Figure~\ref{fig:structure}.
Our general strategy is firstly to map all nodes in the graph to a static vector space we call the node identity space. Then we create a map from a node and a time stamp, via this node identity space, to a node-time space, that is a position in a vector space for each node that changes in time. This second map is trained via the data using an encoder (refer to section~\ref{sec:encoder}) known as temporal graph attention network (TGAT)~\cite{da2020inductive}. The training is via a loss function with two components (refer to section \ref{sec:signals}): contrastive loss and structural matching loss. Contrastive loss is based on a function we call diffusion affinity which measures how close two nodes are in a temporal graph around a given time. Close nodes should be moved together and distant nodes should be separated. Structural matching loss considers a diffusion process centred on nodes at a given time. The distance between two nodes in this loss should be related to how similar their diffusion characters are. 

Training the TGAT to minimise the total loss is performed using structure-aware mini-batches (refer to section~\ref{sec:mini-batch}). For each mini-batch, a centre time is first sampled and a local temporal window is extracted around it. The interactions within this window are aggregated to estimate the local structural clustering pattern, from which several node-time instances are selected as roots. A diffusion-based procedure is then applied around each root to sample a local closely connected nodes. The sampled nodes are mapped to their corresponding node-time instances within the temporal window, and the resulting set of node-time instances are merged to form the mini-batch. Positive and negative pairs are defined within each mini-batch according to the estimated diffusion affinity. Node-time instances with diffusion affinity scores are treated as positive pairs, otherwise, treated as negative pairs. 

A fixed number of mini-batches constitutes one epoch. After each epoch, the TGAT parameters are updated and the embeddings of all node-time instances are recomputed. Training terminates after a fixed number of epochs, after which the learned node-time embedding space is partitioned into communities. The community assigned to a node at a given time is determined by the cluster containing its TGAT embedding at that time.
\subsection{Initialisation}

\subsubsection{Timestamp normalisation}
\label{sec:normalise}
Each temporal interaction is represented as a triple $(u,v,t)$, where $u$ and $v$ denote the interacting nodes and $t$ denotes the timestamp of the interaction. Since raw timestamps can have large numerical values and may vary substantially across datasets, directly using them as model input may make the temporal encoding sensitive to the original time scale. To avoid this issue, we normalize timestamps before feeding them into the model. Specifically, we apply min-max normalisation to map all timestamps into the interval $[0,1]$. Let $t_{\min}$ and $t_{\max}$ denote the minimum and maximum timestamps in the temporal graph. For each timestamp $t$, its normalised value is defined as $\frac{t-t_{\min}}{t_{\max}-t_{\min}}$. After this transformation, the earliest interaction is assigned a normalised timestamp of zero, the latest interaction is assigned a normalised timestamp of one, and all other interactions are placed proportionally between them. We want our method to consider events that are close in time to have more influence on each other. By performing this normalisation the distance between $t_1$ and $t_2$ is now a proportion of the total time of the dataset and we can use $\exp(-|t_2-t_1|/\tau)$ (where $\tau$ is a constant) to control the rate of decay as two events are separated in time without that value depending on the timescale of the data.
\subsubsection{Node features initialisation}
\label{sec:init-node-feat}
For the datasets considered in this study, node attributes are not available. We therefore assign each node a randomly initialised vector~\cite{abboud2020surprising, duong2019node}. These random features provide distinct node identifiers without introducing manually designed structural information, while the TGAT encoder learns temporal and structural representations through neighbourhood aggregation over the observed interactions.

\subsection{Temporal graph encoder}
\label{sec:encoder}
We use a temporal graph attention network (TGAT)~\cite{da2020inductive} as an encoder that maps each node-time instance $i=(u_i,t_i)$ defined by a node identity and a timestamp, into an embedding $\mathbf{z}_{i}$ in the node-time embedding space. In contrast to the original causal TGAT formulation, which aggregates only historical neighbours before the query time, our task is offline dynamic clustering rather than online link prediction. Therefore, the encoder is allowed to use a time-centred local neighbourhood around $t$. For a node-time instance $i=(u_i,t_i)$, we collect interactions incident to $u_i$ from both before and after $t_i$, and select the neighbours whose interaction timestamps are closest to $t_i$. The encoder then applies multi-layer temporal attention to aggregate these local temporal neighbours and outputs the node-time representation $\mathbf{z}_i$, a position in the TGAT's output vector space.

\subsection{Sampling structure-aware mini-batch}
\label{sec:mini-batch}

While some researchers use link prediction tasks to test clustering algorithms, this is often misaligned with community detection, since observed edges may connect nodes from different communities, while nodes in the same community may only be connected through multi-hop paths. Inspired by the modularity-based contrastive formulation of MAGI~\cite{liu2024revisiting}, we define structure-aware training signals using residual structural affinity rather than edge existence. To adapt this idea to temporal networks, we construct a matrix representing node similarity we call the diffusion affinity matrix around a sampled time we call the centre time. This centre time $t_c$ is used to capture the local interaction pattern around a specific time point. Its use is defined in the next sections. It is sampled by choosing an event $(u,v,t)$ at random, and this time becomes the centre time. This strategy means we sample more often in time periods where the data has more events. 

\subsubsection{Diffusion affinity matrix on local aggregated graph}
\label{sec:diffusion}
The aim of this section is to consider the interactions based around a centre time $t_c$ and create a matrix we call the diffusion affinity matrix that represents how ``close'' nodes are at that time.
Given a centre time $t_c$, we aggregate temporal edges within a symmetric window and define
\begin{equation}
    A^{(t_c)}_{uv} = \sum_{(u,v,t)\in E,\ |t-t_c|\le \Delta} \exp\left(-\frac{|t-t_c|}{\tau}\right),
\end{equation}
where $\Delta$ is the window size and $\tau$ controls temporal decay. Let $P^{(t_c)} = (D^{(t_c)})^{-1} A^{(t_c)}$, where $D^{(t_c)}$ is the diagonal weighted degree matrix with $D^{(t_c)}_{ii} = \sum_j A^{(t_c)}_{ij}$. The matrix $P^{(t_c)}$ is therefore the first-order transition probability matrix on the local graph centred at $t_c$. To incorporate high-order structural proximity, we further define the $\ell$-order diffusion affinity matrix as
\begin{equation}
    \Phi^{(t_c)}_{\ell} = \sum_{j=1}^{\ell} D^{(t_c)}(P^{(t_c)})^{j}.
\end{equation}
Therefore, $\Phi^{(t_c)}_{\ell}$ combines direct adjacency and multi-step diffusion information, providing a local high-order structural affinity matrix for the temporal neighbourhood around $t_c$. The intent of this matrix is to capture the strength of interaction between any two nodes, including $\ell$ step interactions, at a given point in time.

Given the local high-order diffusion matrix $\Phi_{\ell}^{(t_c)}$, we define the high-order strength of node $u$ as $s_u=\sum_v \Phi_{\ell,uv}^{t_c}$. This quantity measures the total diffusion affinity from node $u$ within the local temporal neighbourhood centred at $t_c$. We then construct a degree-corrected null model that accounts for the fact that highly connected nodes naturally tend to exhibit larger diffusion affinities. Under this null model, the expected affinity between nodes $u$ and $v$ is given by ${s_u^{(t_c)}s_v^{(t_c)}}/{\sum_w s_w^{(t_c)}}.$ We then define the local high-order affinity matrix as 
\begin{equation}
B_{uv}^{(t_c)}=\Phi_{\ell,uv}^{(t_c)}-\frac{s_u^{(t_c)}s_v^{(t_c)}}{\sum_w s_w^{(t_c)}}.
\end{equation}
A positive value of $B_{uv}^{(t_c)}$ indicates that the observed high-order diffusion affinity between $u$ and $v$ is stronger than expected from their marginal high-order strength, while a negative value indicates weaker than expected affinity. 

Based on the local high-order affinity matrix $B^{(t_c)}$, we first sample a set of root nodes according to their high-order strengths: $p(r=u\mid t_c)={s_u^{(t_c)}}/{\sum_w s_w^{(t_c)}}$, the number of sampled root nodes is denoted by $N_r$. This sampling strategy gives higher probability to nodes with stronger local high-order activity, while still allowing different regions of the local graph to be selected across mini-batches. For each sampled root node $r$, we use the $r$-th row of $B^{(t_c)}$ to identify nodes that have stronger-than-expected high-order affinity with $r$. Specifically, the candidate set associated with root $r$ is defined as $\{v: B^{(t_c)}_{rv}>0\}$. A positive residual value $B^{(t_c)}_{rv}>0$ indicates that the observed high-order diffusion affinity between $r$ and $v$ is larger than expected under the degree-corrected strength null model. Therefore, nodes in ${v: B^{(t_c)}_{rv}>0}$ can be interpreted as structurally associated nodes of $r$ at time $t_c$.

\subsubsection{Sampling node-time instances from nodes}
\label{sec:sampling}

The next step is to use the identified neighbours from the previous section to create a sampled mini-batch of node-time instances for training the TGAT. For each root node $r$, we sample nodes from $\{v: B^{(t_c)}_{rv}>0\}$. Let $\mathcal S_r^{(t_c)}\subseteq \{v: B^{(t_c)}_{rv}>0\}$ denote the sampled positive-residual neighbourhood of root $r$. Since $\mathcal S_r^{(t_c)}$ is obtained from the local aggregated graph around the centre time $t_c$, it is still a set of nodes rather than node-time instances. We therefore map each sampled node $v\in \mathcal S_r^{(t_c)}$ back to its observed node-time instances within the local temporal window. 

For each sampled node $v$, let $\mathcal T_v^{(t_c)} = \left\{ t \mid (v,t)\in\mathcal V,\ |t-t_c|\le \Delta \right\}$. For each timestamp $t\in \mathcal T_v^{(t_c)}$, we define its temporal proximity score as $\pi_{v,t}^{(t_c)} = \exp(\frac{-|t-t_c|}{\tau})$. Since $v$ may be in several events within the window centred on $t_c$, we sample a subset of timestamps with probability proportional to $\pi_{v,t}^{(t_c)}$ which creates a series of samples within the temporal window around $t_c$. This sampling is done without replacement so a single event is never sampled twice. The maximum number of samples is $N_s$. We denote the sampled timestamp set of node $v$ by $\widetilde{\mathcal T}_v^{(t_c)}\subseteq \mathcal T_v^{(t_c)}$. The corresponding sampled node-time instance set is defined as $\mathcal I_v^{(t_c)}=\left\{(v,t)\;\middle|\;t\in\widetilde{\mathcal T}_v^{(t_c)}\right\}$. Thus, each sampled node $v$ is mapped to multiple observed node-time instances around the centre time $t_c$. Given a set of sampled root nodes $\mathcal R^{(t_c)}$, the final mini-batch is constructed as 
\begin{equation}
    \mathcal B^{(t_c)} = \bigcup_{r\in\mathcal R^{(t_c)}} \bigcup_{v\in \{r\}\cup \mathcal S_r^{(t_c)}} \mathcal I_v^{(t_c)}.
\end{equation}
After obtaining the mini-batch, then, the positive sets for node-time instance $i=(u_i,t_i)$ are defined as $\mathcal M_i^+ = \left\{ j\in\mathcal B^{(t_c)}\setminus\{i\} \mid  B_{ij}^{(t_c)} > 0 \right\}$, and the negative set is $\mathcal M_i^- = \left\{ j\in\mathcal B^{(t_c)}\setminus\{i\} \mid  B_{ij}^{(t_c)} \le 0 \right\}$.

\subsection{Training signals}
\label{sec:signals}

\subsubsection{Diffusion-guided contrastive learning}

Based on the positive and negative pairs identified from the diffusion affinity matrix, we employ a softmax-based SimCLR contrastive learning objective~\cite{liu2024revisiting, chen2020simple, wang2021understanding}, that is to say a loss function that pulls the embeddings of positive pairs closer together while pushing those of negative pairs further apart within the same mini-batch. For each node-time instance, the objective assigns a high relative similarity to its positive instances compared with all other instances in the mini-batch. In practical terms, it increases the similarity between node-time instances connected by high diffusion affinity and decreases their similarity relative to instances with low or negative diffusion affinity. Let $\mathbf z_i$ denote the position of node-time instance $i=(u_i,t_i)$ in the node-time space, and we use the cosine similarity $\operatorname{sim}(\mathbf z_i,\mathbf z_j) = \frac{\mathbf z_i^{\top}\mathbf z_j}{\lVert\mathbf{z}_i\rVert \lVert\mathbf{z}_j\rVert}$, where $\lVert \cdot \rVert$ is the Euclidean L2 norm. We define the weighted contrastive loss for node-time instance $i$ as
\begin{equation}
    \mathcal L_{\mathrm{CL}} = - \frac{1}{|\mathcal{B}^{(t_c)}|}\sum_{i\in \mathcal{B}^{(t_c)}} \sum_{j\in\mathcal M_i^+} \omega_{ij}^{(t_c)} \log \frac{ \exp\left( \operatorname{sim}(\mathbf z_i,\mathbf z_j)/\gamma \right) }{\sum_{k\in \mathcal B^{(t_c)}\setminus\{i\}}\exp\left( \operatorname{sim}(\mathbf z_i,\mathbf z_k)/\gamma \right) },
    \label{eqn:clrloss}
\end{equation}
where $\omega_{ij}^{(t_c)} = \frac{ \left| B_{ij}^{(t_c)}\right| }{ \sum_{k\in\mathcal M_i^+} \left| B_{ik}^{(t_c)}\right|}$ is the normalised positive-pair weight, and $\gamma$ is the temperature parameter.

\subsubsection{Structure matching}
Given the local high-order diffusion affinity matrix $\Phi^{(t_c)}_{\ell}$, for each node-time instance $i=(u_i,t_i)$, the diffusion distribution over the nodes in the local graph centred at $t_c$ is $\mathbf{p}_i^{(t_c)}=\frac{\Phi^{(t_c)}_{\ell,u_i v}} {\displaystyle\sum_w \Phi^{(t_c)}_{\ell,u_i w}}$. This distribution serves as a structural signature that summarises the high-order diffusion affinities from the underlying node $u_i$ to all nodes in the local graph. In the structure-matching objective, pairwise similarity between these signatures are used to quantify the structural similarity between node-time instances and guide their representations in the embedding space. Directly computing pairwise similarities (or distances) between these high-dimensional diffusion distributions is computationally expensive, therefore, we apply CountSketch~\cite{charikar2002finding} to obtain a low-dimensional representation of the diffusion distribution. We denote the low-dimensional representation of node-time instance by $\mathbf y_i=\operatorname{CS}\!\left(\mathbf p_i^{(t_c)}\right)$, where $\operatorname{CS}(\cdot)$ denotes the CountSketch projection. Here, $\mathbf{p}_i^{(t_c)}$ is indexed by the lifted node-time instance $i$, but its value is determined by the underlying node $u_i$ and the centre time $t_c$. Therefore, if the same node $u_i$ is lifted to multiple observed times around $t_c$, these node-time instances share the same diffusion distribution $p_i^{(t_c)}$, while their contributions to the training objective are modified by the sampling weights $\alpha_i^{(t_c)}=\alpha_{u_i,t_i}^{(t_c)}$.

We then compare pairwise similarities in two spaces. In the TGAT embedding space, the similarity between two node-time instances $i=(u_i,t_i)$ and $j=(u_j,t_j)$ is defined as  $S_{ij}^{z}=\frac{\mathbf z_i^\top \mathbf z_j}{\lVert \mathbf z_i\rVert\lVert \mathbf z_j\rVert}$, where $\mathbf z_i$ and $\mathbf z_j$ are their learned temporal representations. In the diffusion CountSketch space, the corresponding similarity is defined as  $S_{ij}^{y}=\mathbf y_i^\top \mathbf y_j$. The first similarity measures proximity in the learned representation space, while the second provides a compressed estimate of similarity between high-order diffusion distributions. We only align similarities over non-diagonal pairs in the mini-batch: $\mathcal P=\{(i,j): i\ne j,\ i,j\in\mathcal B^{(t_c)}\}$. Because the TGAT embedding space and the diffusion CountSketch space may have different scales, we standardise the pairwise similarities within each mini-batch before alignment. Specifically, we compute $\mu_z^S=\frac{1}{|\mathcal P|}\sum_{(i,j)\in\mathcal P}S_{ij}^{z}$, and $\sigma_z^S=\sqrt{\frac{1}{|\mathcal P|}\sum_{(i,j)\in\mathcal P}(S_{ij}^{z}-\mu_z^S)^2}$, and define the standardised embedding-space similarity as $\widetilde S_{ij}^{z}=\frac{S_{ij}^{z}-\mu_z^S}{\sigma_z^S}$. Similarly, for the diffusion CountSketch similarities, we compute $\mu_y^S=\frac{1}{|\mathcal P|}\sum_{(i,j)\in\mathcal P}S_{ij}^{y}$, $\sigma_y^S=\sqrt{\frac{1}{|\mathcal P|}\sum_{(i,j)\in\mathcal P}(S_{ij}^{y}-\mu_y^S)^2}$, and obtain $\widetilde S_{ij}^{y}=\frac{S_{ij}^{y}-\mu_y^S}{\sigma_y^S}$. Finally, the pairwise similarity alignment loss is then defined as 
\begin{equation}
    \mathcal L_{\mathrm{matching}}=\frac{1}{|\mathcal P|}\sum_{(i,j)\in\mathcal P}\alpha_i^{(t_c)}\alpha_{j}^{(t_c)}(\widetilde S_{ij}^{z}-\widetilde S_{ij}^{y})^2,
    \label{eqn:simloss}
\end{equation}
where
$$
\alpha_i^{(t_c)} = \frac{ \pi_{u_i,t_i}^{(t_c)} }{ \sum_{t'\in T} \pi_{u_i,t'}^{(t_c)} }, 
$$
and $T$ is the set of sampled times.

By combining \eqref{eqn:clrloss} and \eqref{eqn:simloss}, the total loss is $\mathcal{L}_{\text{tot}}=\mathcal{L}_{\text{CL}}+\mathcal{L}_{\text{matching}}$. We use the total loss to update the TGAT which maps from the node identity space and a time to a position in our node-time space. Effectively this means we can take a node-time pair and turn it into a position in a vector space. The TGAT training is iterated for a fixed number of epochs. 
\subsection{Clustering and creation of communities}
\label{sec:clustering}
After training, the TGAT encoder provides an embedding vector $\mathbf{z}_i$ for each node-time instance $i=(u_i,t_i)$. These embeddings capture the local temporal interaction patterns and structural relationships learned from the diffusion-guided contrastive objective.

To obtain dynamic communities, we apply a clustering algorithm to the set of node-time embeddings $\mathcal Z=\{\mathbf z_{i}\mid (u_i,t_i)\in\mathcal V\}$, where $\mathcal V$ denotes the set of observed node-time instances. In this work, we employ K-means to cluster on the node-time embedding space. The number of clusters is denoted by $K$.

The clustering procedure assigns each node-time instance $(u,t)$ to a community label $c_{u,t}\in\{1,\ldots,K\}$. Node-time instances with similar temporal interaction patterns and diffusion structures are therefore grouped into the same community. The resulting labels define the dynamic community structure of the temporal graph, allowing a node to belong to different communities at different time points.
\section{Datasets}
A problem with dynamic community algorithms in temporal graphs is that it is difficult to find datasets with ground-truth community/clustering labels. These limitations mainly come from: (1) the fact that most temporal graph datasets do not contain community/cluster labels of nodes; and (2) many datasets are designed for binary classification of nodes (e.g., risky users), and these datasets are not suitable for studying dynamic community detection. Because reliable node-time-level ground-truth community labels are difficult to obtain in real temporal networks, our quantitative evaluation mainly relies on synthetic temporal networks with controlled dynamic community structure~\cite{bazzi2020framework}. We further use a large-scale OpenAlex collaboration network as a real-world case study to examine scalability and interpretability.
\subsection{Synthetic temporal network with dynamic clusters}
\begin{figure}
    \centering
    \includegraphics[width=0.7\linewidth]{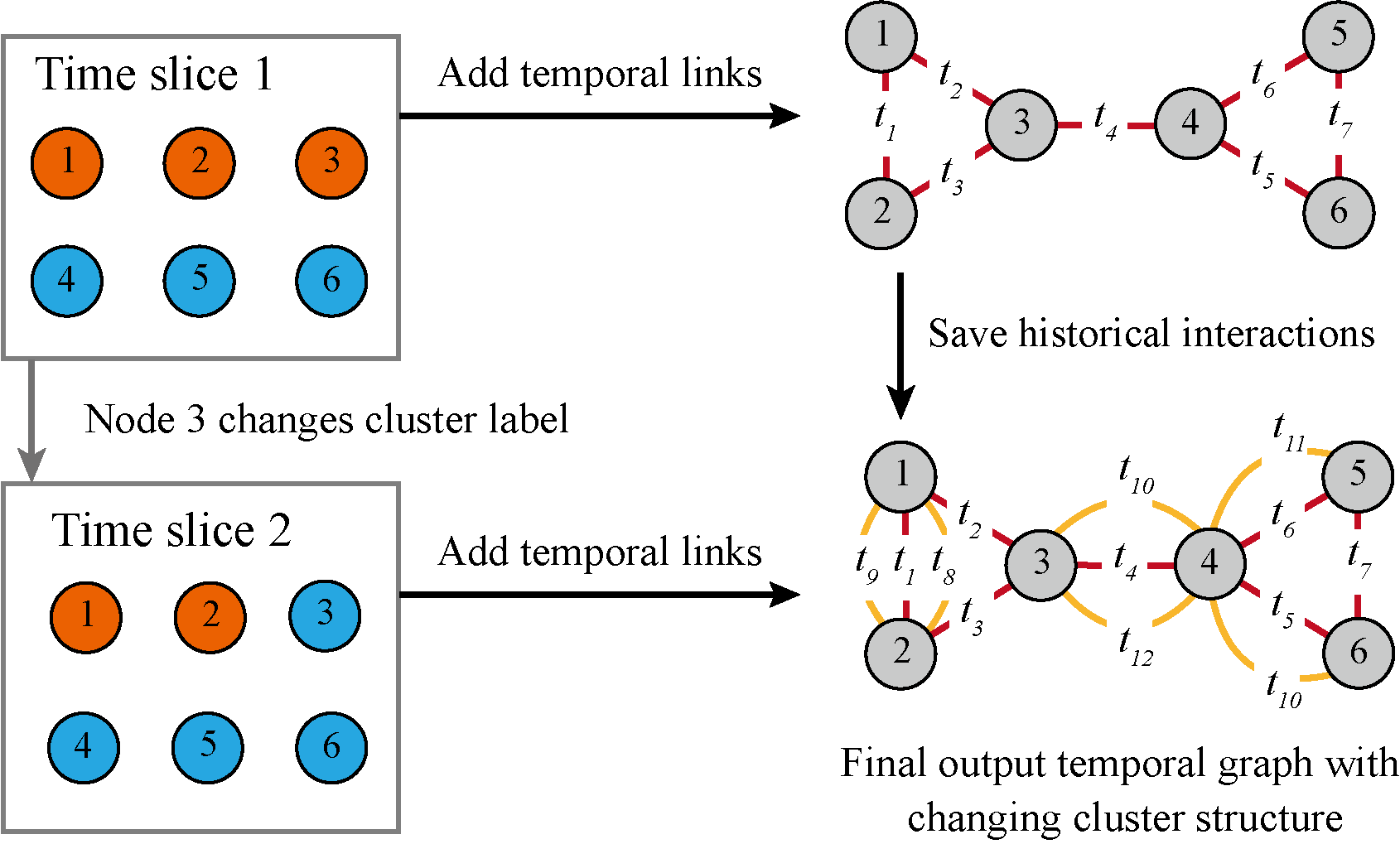}
    \caption{The figure uses two time slices as an example to explain how a temporal graph with dynamic community structure can be synthesised. The left side first generates the community structure of the first time slice, and then samples the timestamp of the timestamped interaction from the Poisson process (annotated with the red line as the interaction in the first time slice). In the second time slice, the community label of node 3 changes. The timestamped interaction (yellow line) in the second time slice is then synthesised in the same way, and inherits all historical interactions.}
    \label{fig:synthetic_net}
\end{figure}
The paper~\cite{bazzi2020framework} proposed a generative framework for synthesising multilayer community structure that evolves across layers. Their method does not directly generate temporal edges with continuous timestamps. Instead, it first produces a multilayer partition, assigning each node a community label in each layer, i.e., $(u,l)\mapsto c_u^{(l)}$. In this framework, the dependency tensor controls how strongly a node's community membership in one layer depends on its membership in other layers, such as whether it tends to retain its previous label, while the null distribution controls how a new label is resampled when such dependence is not enforced. In this way, the model captures meso-scale dynamics across discrete layers, allowing communities to persist, shift, or reorganise over layers.

Building upon this layer-wise partition, we further introduce an event generation procedure to construct a temporal network in the form of a link stream. Specifically, for each layer, all candidate node pairs are first divided into intra-community pairs and inter-community pairs according to the sampled community labels. A mixing parameter $\mu$ is then used to control the proportion of inter-community interactions~\cite{karrer2011stochastic}, while we also introduce the link-node ratio to control the overall scale of interactions within the layer. Given the number of events to be generated, node pairs are sampled with replacement from the corresponding candidate sets, which allows repeated interactions between the same pair of nodes within a layer. To assign continuous timestamps to these sampled interactions, we assume that event arrivals follow a Poisson process: inter-event times are drawn from an exponential distribution and cumulatively summed to obtain a monotonically increasing sequence of timestamps. In this way, the original multilayer community structure, which is defined only over discrete layers, is transformed into a temporal interaction network while reflecting the evolving community organisation across layers. 

\subsection{OpenAlex computer science collaboration network}
In addition to synthetic temporal networks, we also use a real-world scientific collaboration network constructed from OpenAlex\footnote{This dataset is freely available at \url{https://openalex.org}}~\cite{priem2022openalex}. OpenAlex is a large-scale open bibliographic database that provides metadata for scholarly works, authors, publication dates, research topics, and related entities. In this study, we focus on scholarly works published between 2016 and 2025 that are labelled as belonging to the field of computer science. The resulting dataset contains 6,524,460 works. However, the OpenAlex dataset is classified by an LLM so it not an error-free dataset. Inspection of the data indicates that the large majority of publications correspond to recognisable CS research, while a comparatively small number originate from adjacent or potentially unrelated disciplines, such as law and linguistic research. We will further discuss our data-cleaning procedure in section~\ref{sec:openalex}.

We then construct a temporal co-authorship network from the cleaned dataset. Each node represents an author, and each temporal link represents a co-authorship relation observed at the publication time of a paper. In this way, the OpenAlex records are transformed into a link stream. The OpenAlex metadata also provides research topic and subfield information, which is used to interpret the detected communities. In particular, we use major computer science subfields, including Artificial Intelligence, Computer Networks and Systems, Computer Vision and Pattern Recognition, Information Systems, Signal Processing, and other related areas, to characterise the semantic composition of the inferred communities and analyse their temporal evolution. 
\section{Experiments and results}
In this section, we present the experimental evaluation of the proposed method. We first introduce the evaluation metrics for dynamic community detection, then report the performance of community detection algorithms on synthetic networks with known dynamic community labels, then present an ablation study, and runtime scalability. Finally, we provide a case study on a large-scale OpenAlex collaboration network to examine the method’s applicability to real-world temporal data.
\subsection{AMI and ARI in dynamic community setting}
In the static graph setting, AMI (adjusted mutual information)~\cite{vinh2009information, danon2005comparing} is a commonly used metric to measure the performance of a clustering/community detection algorithm. AMI is computed by comparing two community structures over the node set $V$. In our temporal setting, we extend the sample space from nodes to observed node-time pairs. Let $\mathcal{V}=\{(u,t)\in V\times T \ |\ \exists \ v \text{ s.t. } (u,v,t)\in E\}$ be the set of observed node-time pairs, and let $N=|\mathcal{V}|$. We denote by $\mathcal{C}^{\text{gt}}(u,t)$ the community labels assigned to $(u,t)\in \mathcal{V}$ and $\mathcal{C}^{\text{pred}}(u,t)$ the community labels assigned to $(u,t)\in \mathcal{V}$ by the ground-truth community structure and the predicted community structure, respectively. Let $n^{\text{gt}}_r$ be the number of node-time pairs assigned to community $r$ in $\mathcal{C}^{\text{gt}}$, $n_s^{\text{pred}}$ the number assigned to community $s$ in $\mathcal{C}^{\text{pred}}$, and $n_{r,s}^{\text{gt, pred}}$ the number assigned to community $r$ in $\mathcal{C}^{\text{gt}}$ and community $s$ in $\mathcal{C}^{\text{pred}}$. The mutual information is then computed as 
\begin{equation}
    I(\mathcal{C}^{\text{gt}},\mathcal{C}^{\text{pred}})=\sum_r\sum_s\frac{n_{r,s}^{\text{gt, pred}}}{N}\log\left(\frac{Nn_{r,s}^{\text{gt, pred}}}{n^{\text{gt}}_rn_s^{\text{pred}}}\right)
\end{equation}
The metric measures the agreement between the two partitions. To eliminate the effect of random agreement, AMI corrects mutual information by subtracting a random partition baseline. Let $\mathbb{E}[I(\mathcal{C}^{\text{gt}},\mathcal{C}^{\text{pred}})]$ denote the expected mutual information under a random baseline. This baseline is defined under the permutation model, where the community-size marginals $\{n_r^{\text{gt}}\}$ and $\{n_s^{\text{pred}}\}$ are fixed, while the assignments are assumed to be random. Equivalently, the contingency table entries $n_{r,s}^{\text{gt, pred}}$ are treated as random variables subject to these fixed marginals. The adjusted mutual information is then defined as 
\begin{equation}
\operatorname{AMI}=\frac{I(\mathcal{C}^{\text{gt}},\mathcal{C}^{\text{pred}})-\mathbb{E}[I(\mathcal{C}^{\text{gt}},\mathcal{C}^{\text{pred}})]}{\frac{1}{2}(H(\mathcal{C}^{\text{gt}})+H(\mathcal{C}^{\text{pred}})-\mathbb{E}[I(\mathcal{C}^{\text{gt}},\mathcal{C}^{\text{pred}})])}
\end{equation}

Similarly, the adjusted Rand index (ARI) can be extended to the dynamic community setting by treating each observed node-time pair $(u,t)\in\mathcal{V}$ as one sample. Rather than comparing community labels directly, ARI evaluates the agreement between two community structures in terms of pairwise assignments over $\mathcal{V}$. For any two distinct node-time pairs $x,y\in\mathcal{V}$, one considers whether $x$ and $y$ are assigned to the same community or to different communities under $\mathcal{C}^{\text{gt}}$ and $\mathcal{C}^{\text{pred}}$. Let $a$ be the number of pairs that are placed in the same community in both partitions, and let $b$ be the number of pairs that are placed in different communities in both partitions. The Rand index~\cite{rand1971objective} is then defined as
\begin{equation}
\operatorname{RI}=\frac{a+b}{\binom{N}{2}},
\end{equation}
where ${\binom{N}{2}}$ is the number of possible pairs for $N$ node-time instances.
Since a non-negligible agreement may arise by chance, ARI~\cite{hubert1985comparing} adjusts RI under the same permutation model used for clustering comparison. Using the contingency table $\{n_{r,s}^{\text{gt,pred}}\}$, ARI can be written as
\begin{equation}
\operatorname{ARI}
=\frac{\sum_{r,s}\binom{n_{r,s}^{\text{gt,pred}}}{2}-\frac{\left(\sum_r\binom{n_r^{\text{gt}}}{2}\right)\left(\sum_s\binom{n_s^{\text{pred}}}{2}\right)}{\binom{N}{2}}}{\frac{1}{2}\left[\sum_r \binom{n_r^{\text{gt}}}{2}+\sum_s \binom{n_s^{\text{pred}}}{2}\right]-\frac{\left(\sum_r \binom{n_r^{\text{gt}}}{2}\right)\left(\sum_s\binom{n_s^{\text{pred}}}{2}\right)}{\binom{N}{2}}}.
\end{equation}
Therefore, in the dynamic setting, ARI measures the extent to which the predicted community structure preserves the pairwise grouping relations among observed node-time pairs, after correcting for chance agreement.

We do not use classification-based metrics such as accuracy or F1, because the label given to a community is arbitrary; only how nodes are divided is relevant. A partition of four nodes into communities $(0,0,1,1)$ is the same community structure as $(1,1,0,0)$. In this case, AMI and ARI correctly return the highest score because they compare partitions rather than raw label identities. In contrast, accuracy or F1 would incorrectly penalise such a result, since none of the predicted labels matches the ground-truth label values directly.

\subsection{Performance on synthetic temporal graphs}
\begin{figure}[t]
    \centering
    \includegraphics[width=\linewidth]{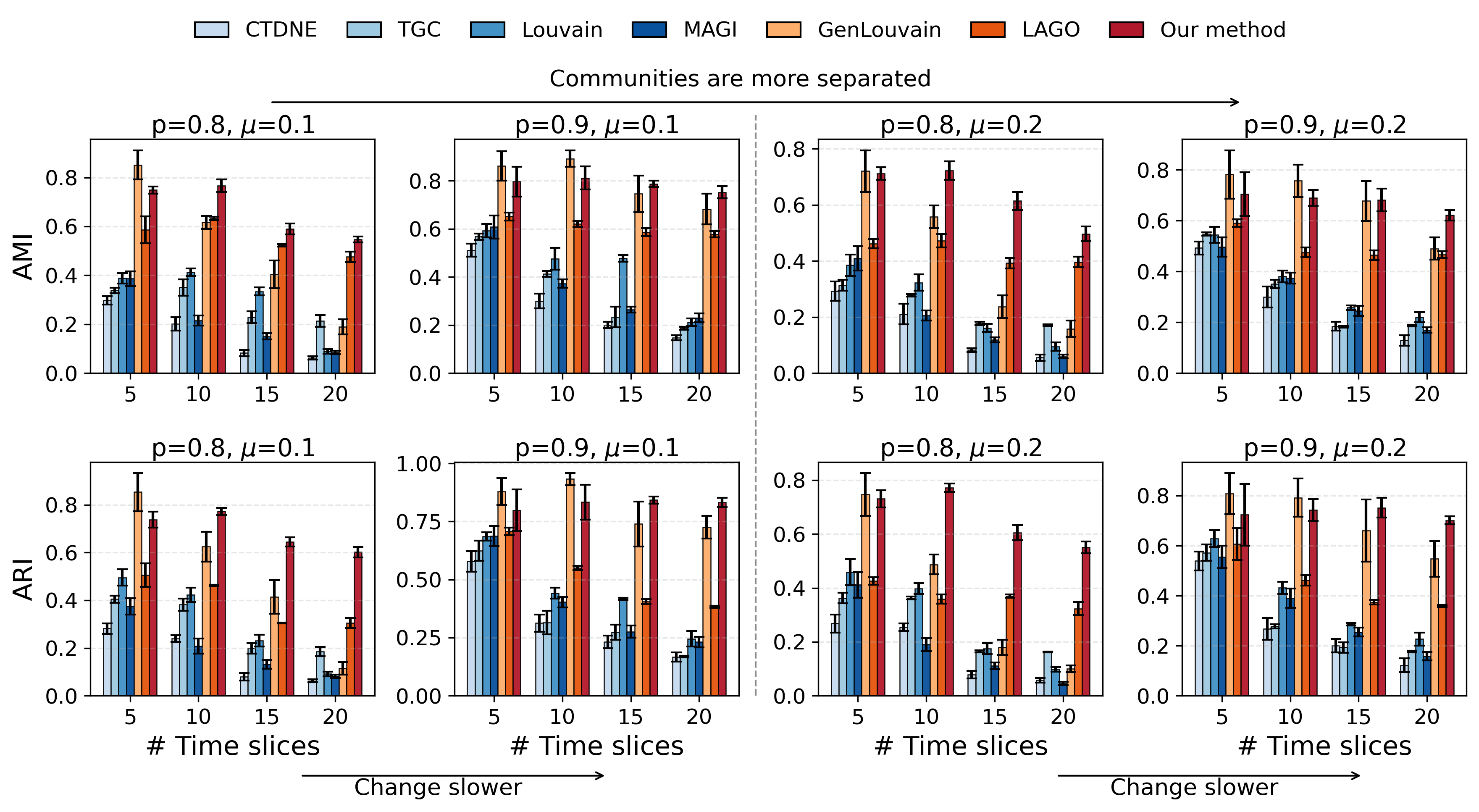}
    \caption{AMI and ARI comparison across different synthetic temporal network settings. The top row reports AMI and the bottom row reports ARI under different combinations of community persistence probability $p$ and mixing parameter $\mu$. The experimental data were obtained through ten repeated experiments, and the error bar represents the standard error of the mean for the repeated experiments. The x-axis denotes the number of time slices. Static representation-based methods, including CTDNE, TGC, and Louvain, generally obtain lower scores, especially as the number of slices increases. Dynamic community detection methods, including GenLouvain, LAGO, and the proposed method, generally achieve higher scores because they allow node memberships to vary over time.}
    \label{fig:ami-ari}
\end{figure}

The experiment compared the performance of the static community detection methods and the dynamic community detection methods on synthetic data. The static baselines comprise Louvain~\cite{blondel2008fast}, which maximises modularity on the graph aggregated over the full observation period; CTDNE~\cite{nguyen2018continuous}+K-Means, which learns node embeddings from time-respecting random walks before clustering them with K-Means; TGC~\cite{Liu2025DeepDatasets}, which combines temporal graph representation learning with clustering but assigns a single global cluster label to each node; and MAGI~\cite{liu2024revisiting}, a static deep graph clustering method that formulates modularity maximisation from a contrastive-learning perspective. Among the dynamic baselines, GenLouvain~\cite{bazzi2016community} discretises the temporal graph into multilayer snapshots and greedily optimises multilayer modularity, whereas LAGO~\cite{brabant2025discovering} directly optimises longitudinal modularity~\cite{brabant2025longitudinal} over node-time assignments on the continuous-time link stream.

Fig.~\ref{fig:ami-ari} reports the AMI and ARI results on synthetic temporal networks under different temporal stability and community separation settings. Compared with the dynamic community detection methods, the static community detection baselines, including CTDNE, TGC, and Louvain, generally show weaker performance on both AMI and ARI. This gap becomes more evident when the community structure changes more frequently over time. Since these static baselines assign one fixed community label or representation to each node, they are less able to capture cases where the same node belongs to different communities at different timestamps. In contrast, dynamic methods such as LAGO, GenLouvain, and our method explicitly model temporal variation in community membership, and therefore better fit the node-time evaluation setting. Another finding across both AMI and ARI is that GenLouvain performs competitively when the number of layers is small, but its performance tends to decrease as the number of layers increases. In this synthetic setting, it means that the community evolution process contains more temporal stages, and nodes have more opportunities to change their community labels across layers. Therefore, the community detection task becomes more challenging because the method needs to correctly track community assignments over a longer sequence of changes. It is also worth noting that, in this experiment, the multilayer network used as the input of GenLouvain is constructed strictly according to the same layer boundaries used during synthetic data generation. This gives GenLouvain favourable prior information about the temporal segmentation of the data. However, in real temporal networks, such clear and ground-truth-aligned temporal boundaries are usually unavailable. Therefore, its performance in this setting may overestimate its effectiveness in practical link stream analysis. In contrast, our method does not rely on predefined layer boundaries and remains more stable as the number of layers increases, showing its advantage in recovering dynamic community structures over longer and more frequently changing temporal processes.

\begin{figure}
    \centering
    \includegraphics[width=0.8\linewidth]{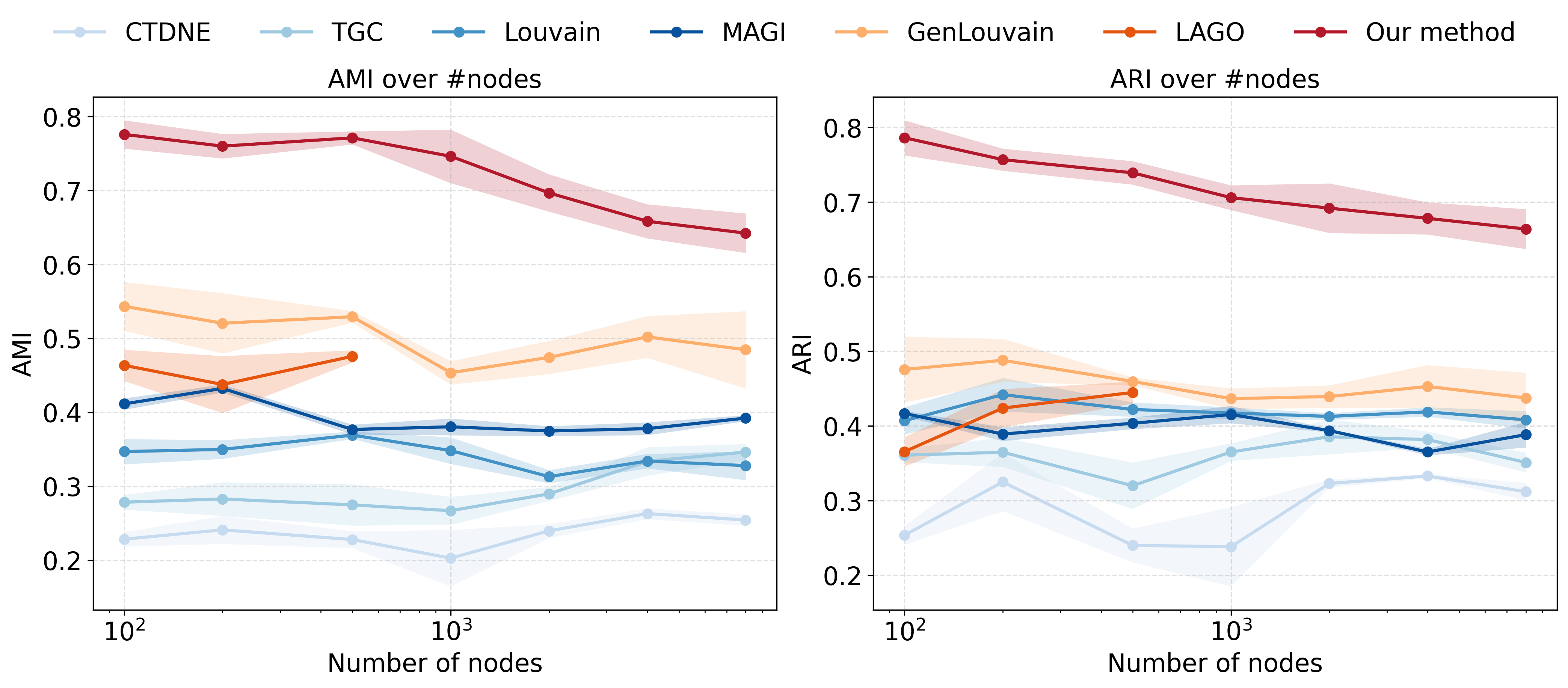}
    \caption{Performance comparison under different network scales. AMI and ARI scores of different community detection methods on synthetic dynamic networks with varying numbers of nodes. The networks are generated with $p=0.8$, $\mu=0.2$, and 10 time slices. The data points were averaged after 10 repetitions. The shaded area represents the standard error of the mean for this set of experiments. Missing values correspond to cases where the algorithm exceeded the 60-minute runtime limit and was terminated.}
    \label{fig:ami-ari-nodes}
\end{figure}

We further examine the effect of network scale on algorithm performance in Fig.~\ref{fig:ami-ari-nodes}. The synthetic networks are generated under the setting $p=0.8$, $\mu=0.2$, and 10 time slices, with different numbers of nodes. We then compare the AMI and ARI scores of different methods. A 60-minute runtime limit is imposed on all algorithms; therefore, the missing results of LAGO at larger scales indicate that the corresponding runs were terminated due to timeout. Overall, as the number of nodes increases, most methods show a slight decline or fluctuation in both AMI and ARI. This suggests that temporal networks make dynamic community recovery more challenging. One possible reason is that increasing the number of nodes also increases the number of node-time assignments to be inferred, requiring the algorithms to recover both the community structure and its temporal evolution in a larger search space.

Nevertheless, dynamic community detection methods and dynamic clustering methods generally outperform static or node-level representation methods. Static methods usually aggregate or weaken temporal information, making it difficult to capture changes in node community membership over time. In contrast, dynamic methods explicitly model dynamic community structure and therefore maintain higher AMI and ARI scores across different network sizes. The proposed method achieves the best or near-best performance on both metrics, with relatively small fluctuations as the network size increases. This indicates that learning time-node level representations remains stable and effective on larger dynamic networks.
\subsection{Ablation study}
\begin{figure}
    \centering
    \includegraphics[width=\linewidth]{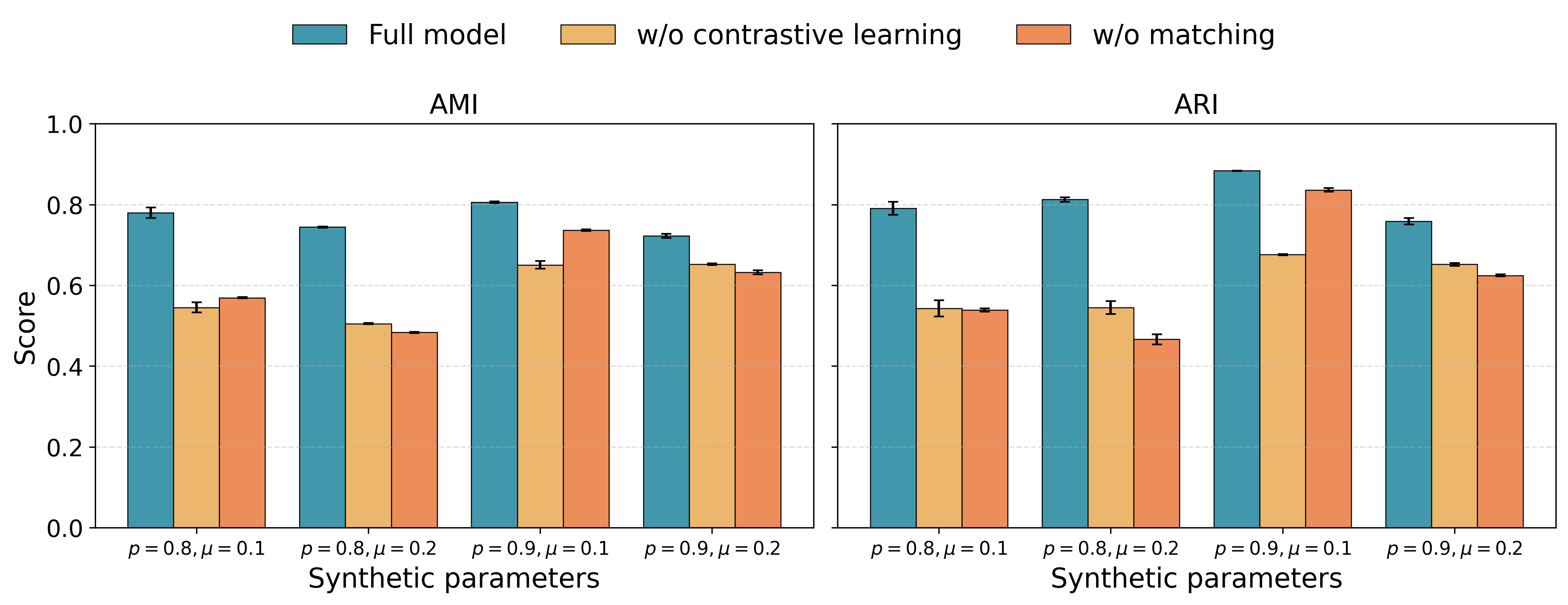}
    \caption{Ablation study of the proposed model under a set of parameters where the number of time slices is fixed as $l=10$. The figure reports mean AMI and ARI scores with standard error of the mean (SEM) over 10 repeats.}
    \label{fig:ablation}
\end{figure}
To assess the contributions of the two training signals considered in this ablation, we compare the full model with two variants: one without the diffusion-guided contrastive-learning objective (\textit{w/o contrastive learning}) and one without the structure-matching objective (\textit{w/o matching}). Figure~\ref{fig:ablation} reports their AMI and ARI scores under four combinations of the community persistence probability $p$, at which nodes keep their community lablels over time, and mixing parameter $\mu$, which controls the proportion of links within/between communities. The full model achieves the highest mean AMI and ARI in every setting, while removing either objective consistently reduces performance. The relatively small error bars further indicate that these trends are stable across repeated runs.

The magnitude of the reduction varies across the synthetic settings. Both ablations generally incur larger performance losses when $p=0.8$, suggesting that the two objectives become particularly important when community memberships change more frequently. Their relative contributions nevertheless depend on the graph configuration. When $p=0.9$ and $\mu=0.1$, removing contrastive learning causes a considerably larger reduction than removing matching. In contrast, when $p=0.8$ and $\mu=0.2$, removing matching produces the lowest scores, particularly for ARI. In the remaining settings, the effects of the two ablations are more comparable, although removing matching gives slightly lower mean scores when $p=0.9$ and $\mu=0.2$.

These results suggest that neither objective uniformly dominates across all settings. Instead, they provide complementary forms of supervision: the contrastive-learning objective organises the embedding space using positive and negative node-time pairs derived from diffusion affinity, whereas the matching objective aligns pairwise embedding distances with differences in local diffusion structure. Combining both signals therefore produces more robust dynamic community assignments across different levels of temporal persistence and community separation.
\subsection{Runtime analysis}
\begin{figure}
    \centering
    \includegraphics[width=\linewidth]{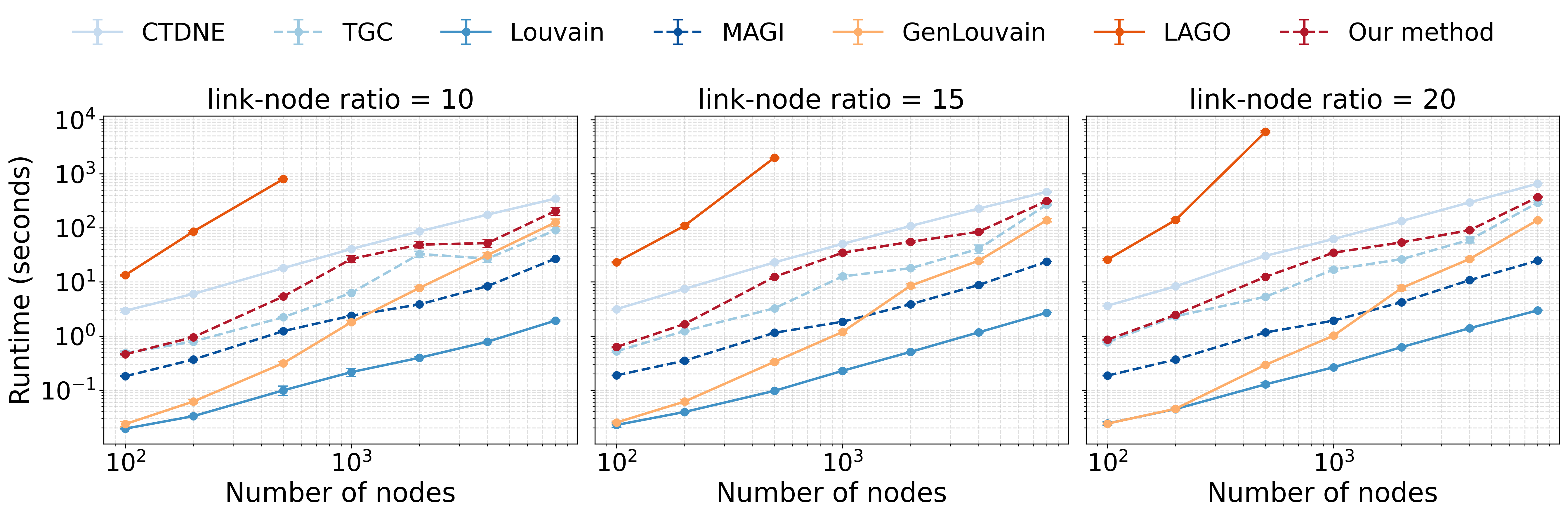}
    \caption{Runtime comparison across different network scales. The density of the synthesised temporal graph increases as the link-node ratio increases, meaning that with the same number of nodes, there are more temporal links. The x-axis shows the number of nodes, while the y-axis reports runtime in seconds. Both axes are plotted on a logarithmic scale. Error bars indicate the standard error of the mean across ten repeated runs. For algorithms based on neural networks, we record the run time of thirty epochs; for non-iterative algorithms, we report the runtime of one complete run. CPU-based algorithms were evaluated on an Intel Xeon Gold 6142 CPU @ 2.60GHz, while GPU-based models were evaluated on an NVIDIA A100 GPU, and their running results are annotated with dashed lines. Runs exceeding 60 minutes were treated as timed out and are not shown.}
    \label{fig:runtime}
\end{figure}
In the previous sections, we listed and analysed the theoretical time complexity of each algorithm. In this section, we further examine their empirical running time in practical settings. It should be noted that the implementations of some algorithms, such as Louvain and LAGO, are not readily amenable to hardware acceleration. These methods are therefore evaluated on a CPU-based platform equipped with an Intel Xeon Gold 6142 processor. In contrast, neural network-based methods can benefit from GPU acceleration. They were evaluated on an NVIDIA A100 GPU~\cite{king2017apocrita}, and their running results are annotated with dashed lines. It is hard to fairly compare methods that use GPU with methods that use CPU, but those CPU based methods had no GPU implementation available. The rate of increase, however, remains a useful point of comparison. 

Fig.~\ref{fig:runtime} compares the empirical runtime of different methods as the network scale increases. Overall, the static or snapshot-based methods show relatively low runtime, but they do not explicitly model node-time level dynamic community assignments. In contrast, methods designed for dynamic community detection usually require additional computation to capture temporal evolution, which leads to higher runtime in large temporal networks. 

Static methods such as Louvain are highly efficient because they operate on an aggregated graph, but this comes at the cost of discarding temporal information. CTDNE preserves temporal ordering through temporal random walks, while the additional walk sampling and Skip-gram training make its runtime grow substantially with network size. For dynamic community detection methods, modelling temporal community evolution introduces further complexity. LAGO grows rapidly and becomes infeasible at larger scales due to the cost of longitudinal modularity optimisation. GenLouvain reduces the cost by discretising the temporal network into static layers, but this also loses part of the fine-grained temporal information.

Neural-network-based methods reduce runtime by using mini-batch training and local temporal neighbourhood aggregation. TGC achieves stable runtime under this strategy, but it learns a single static representation for each node and does not explicitly model dynamic community assignments. Our method extends this framework to node-time clustering while keeping runtime close to TGC, showing that the dynamic community assumption does not introduce substantial extra computational overhead. Together with the AMI and ARI results, this demonstrates that our method provides a better balance between scalability and dynamic clustering accuracy.

\subsection{Hyper-parameters and tuning}

\begin{table}[t]
    \centering
    \begin{tabular}{l l m{5cm}}
    \toprule
    Description & Notation & Details in section\\
    \midrule
    The number of hops & $\ell$ & section~\ref{sec:diffusion} \\
    The length of temporal context & $\Delta$ & section~\ref{sec:diffusion}\\
    Temporal decay & $\tau$ & section~\ref{sec:diffusion} \\
    Number of sampled centre times & $N_c$ & section~\ref{sec:sampling}\\
    The number of sampled root nodes & $N_r$ & section~\ref{sec:sampling}\\
    The number of sampled node-time instances & $N_s$ & section~\ref{sec:sampling}\\
    The number of clusters & $K$ & section~\ref{sec:clustering} \\
    \bottomrule
    \end{tabular}
    \caption{Model parameters.}
    \label{tab:parameters}
\end{table}
The operation of the proposed method is primarily governed by the temporal window size $\Delta$, the temporal decay parameter $\tau$, the diffusion order $\ell$, the number of sampled node-time instances $N_s$, the number of sampled root nodes $N_r$, and the number of sampled centre-times $N_c$. The parameters $\Delta$, $\tau$, and $\ell$ determine the temporal and structural range of the local diffusion process. The parameters $N_r$ and $N_s$ control the composition and size of the mini-batches, while $N_c$ determines the amount of temporal coverage provided during training. After representation learning, the number of clusters $K$ controls the granularity of the final dynamic partition. These parameters are discussed below in terms of their roles, selection criteria, and effects on clustering performance. The selection of model parameters was carried out to achieve a balance among computational efficiency, scalability, and representation quality.

The temporal window size $\Delta$ determines the interactions available when constructing the local diffusion process. A small $\Delta$ produces sparse local graphs and may leave insufficient active nodes for constructing a valid mini-batch. Conversely, an excessively large window may combine interactions generated under different temporal contexts and thereby weaken the temporal specificity of the resulting representations. $\tau$ is a parameter related to $\Delta$, which determines the contribution of interactions within this window through a weight. These two parameters are determined by the nature of the dataset. The number of sampled node-time instances $N_s$, and the number of sampled centre-time instances $N_c$, affect batch size, and the time points involved in the training. We conducted a sensitivity analysis of these parameters on the synthetic data. $N_s$ specifies how many observed node-time instances are associated with each node selected. Using $N_s=1$ maps each selected node to a single nearby observation and may discard relevant temporal evidence. In contrast, a larger $N_s$ introduces temporally redundant instances, increases the mini-batch size, and may overemphasise frequently active nodes. 

The results show in Fig.~\ref{fig:ami-ari} used a $N_s=2$. We changed the value of $N_s$ and $N_r$ and compared the dynamic AMI and dynamic ARI between dynamic community structure calculated by our model and the ground-truth value. Compared with this setting, $N_s=1$ reduced AMI and ARI by approximately $4.4\%$ and $4.6\%$, while $N_s=3$ reduced them by approximately $3.0\%$ and $3.8\%$, respectively. The latter setting also increased the training cost because it produced larger mini-batches. The number of sampled centre time $N_c$ determines the amount of temporal coverage provided during training. In the current implementation, each mini-batch is conditioned on one centre time. Increasing $N_c$ improves temporal coverage but will increase training time as well. Doubling the number of sampled centre times from 2,880 to 5,760 improved AMI by only 1.49\% and ARI by 0.50\%, while increasing training time by $\sim112\%$. Doubling $N_c$ again to 11,520 increases AMI by only 0.11\%, and ARI by 0.08\%. We therefore use $N_c=2,880$ as the default parameter setting because it provides a more favourable balance between clustering performance and computational cost.
\subsection{Case study on the OpenAlex collaboration network}
\label{sec:openalex}
\begin{figure}
    \centering
    \subcaptionbox{Complementary cumulative distribution function of node degree.\label{fig:CCDF}}{\includegraphics[width=.48\linewidth]{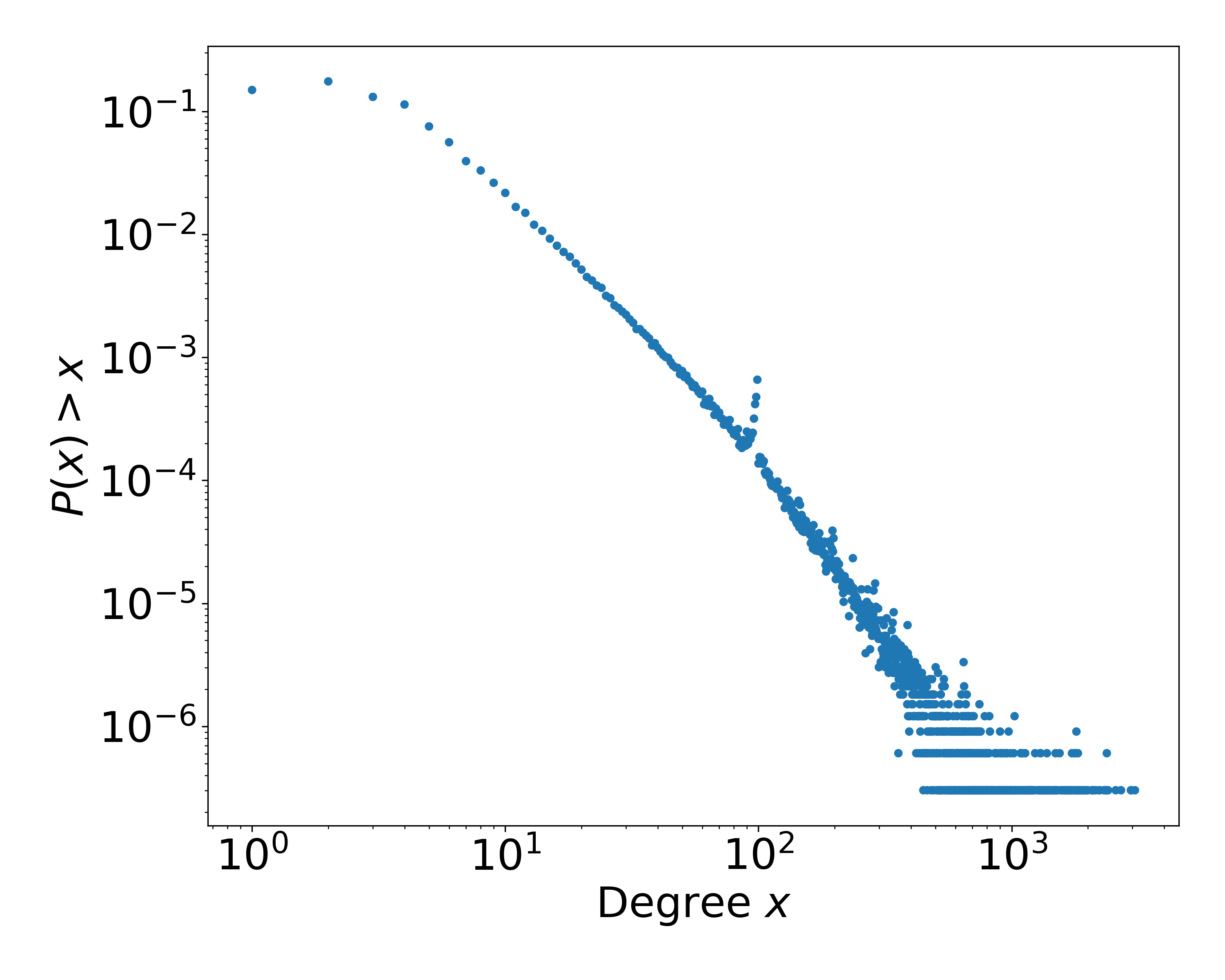}}
    \hspace{10px}
    \subcaptionbox{Number of collaborations per year. \label{fig:collaboration-count}}{\includegraphics[width=.48\linewidth]{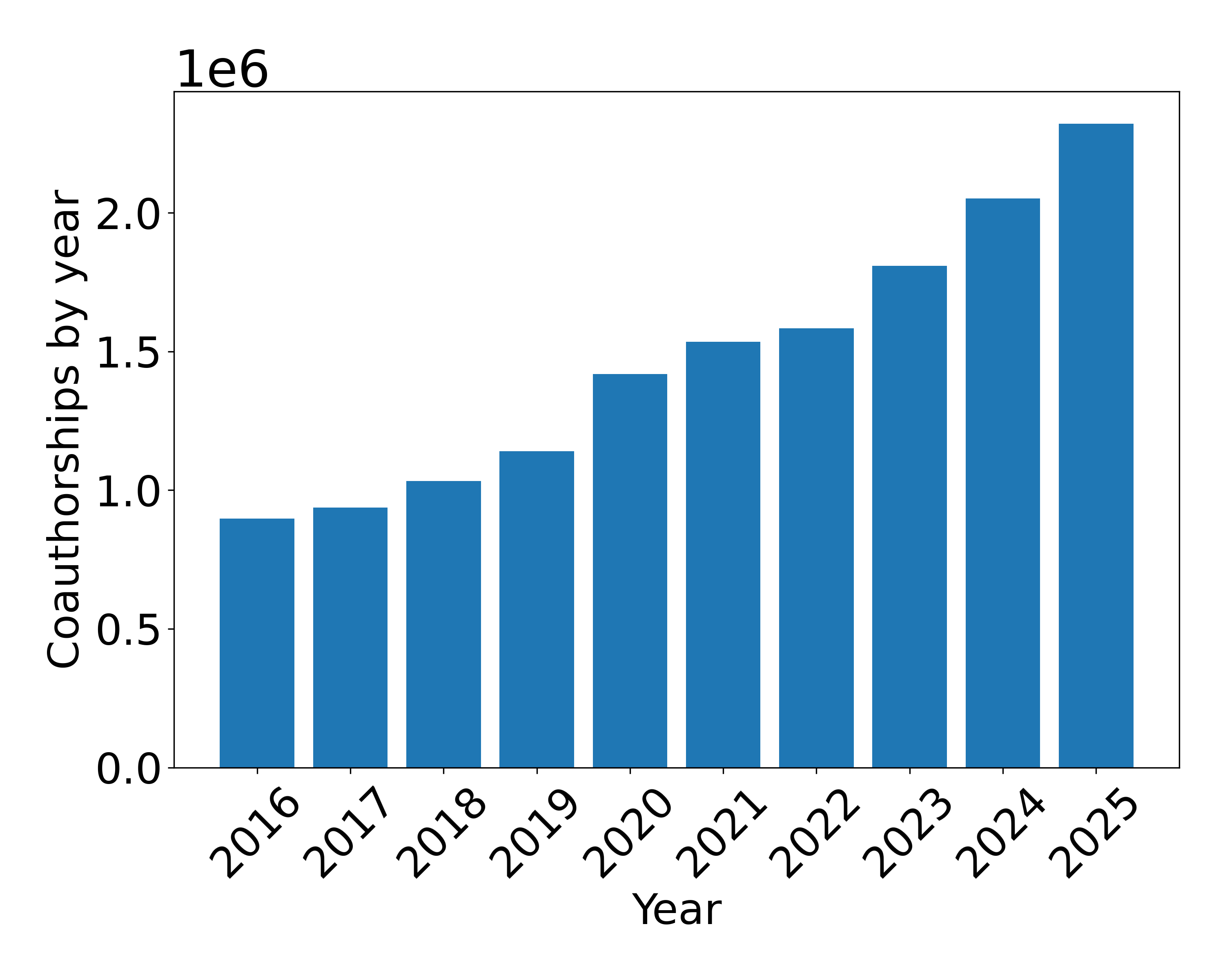}}
    \caption{(a) Complementary cumulative distribution function (CCDF) of node degree on a log-log scale. The red dashed line marks the top 10\% degree threshold. (b) Annual number of temporal collaboration links and the corresponding year-over-year growth rate. The bars show the number of temporal links in each year, while the line indicates the relative increase compared with the previous year.}
    \label{fig:ccdf-and-count}
\end{figure}

To show the potential of dynamic community detection using this method, we take a deep dive into an openly available large-scale dataset of interest to the community that studies temporal networks. Following this perspective, we further apply the proposed model discussed in the previous sections to a real-world large-scale temporal network, with the aim of detecting dynamic community structures and providing empirical analysis on real temporal collaboration data. Specifically, we use the computer science subset of OpenAlex~\cite{priem2022openalex}, focusing on scholarly works published between 2016 and 2025 that are labelled as belonging to the field of computer science. The original dataset contains 6,524,460 works, which it contains datasets, paratexts, etc. We first restrict the data to published articles classified by OpenAlex as belonging to computer science, resulting in 3,303,415 authors and 3,140,693 publications. 
The dataset provides broad coverage of the CS research community, although the OpenAlex classification is not entirely error-free. Inspection of the data indicates that the large majority of publications correspond to recognisable CS research, while a comparatively small number originate from adjacent or potentially unrelated disciplines, such as law and linguistic research. Such cases may arise from classification errors in the underlying source data or from OpenAlex's automated field-classification procedure. In addition, author lists for some publications may be incomplete or truncated (Fig.~\ref{fig:CCDF} shows a jump around the degree of 100), resulting in missing co-authorship relations. These limitations introduce some noise into the network which we seek to mitigate through the filtering procedure described below. In addition, the initial co-authorship network consists of approximately 200,000 small disconnected components, which are less meaningful for community detection. Thus, we are interested in community structure within the main CS research network and therefore restrict the subsequent analysis to its largest connected component (LCC). Nodes belonging to disconnected components cannot be associated with communities in the LCC through the observed co-authorship structure and are consequently outside the scope of this analysis. This restriction reduces the network to 2,295,529 authors ($\sim 70\%$) and to 13,129,037 co-authorship links ($\sim 90\%$). The LCC restriction also provides a secondary data-quality benefit. Inspection of the small disconnected components reveals a disproportionate number of publications that appear to have been incorrectly classified as computer science. In many such cases, the corresponding authors primarily publish outside CS and have only a small number of publications assigned to the CS field. They consequently have few or no co-authorship connections to the main CS research network and tend to occur as isolated nodes or small components. Restricting the analysis to the LCC therefore removes many of these peripheral and potentially misclassified records while retaining the large, densely connected core of the CS co-authorship network. Then we use the cleaned dataset to construct a temporal network for analysis. In addition, we use the subfield information provided by OpenAlex, including Artificial Intelligence, Computer Networks and Systems, Computer Vision and Pattern Recognition, Information Systems, Signal Processing, and other related areas, to interpret the detected communities and analyse their temporal evolution.

\begin{figure}
    \centering
    \subcaptionbox{Elbow plot for selecting the number of clusters.\label{fig:WCSS}}{\includegraphics[width=0.45\linewidth]{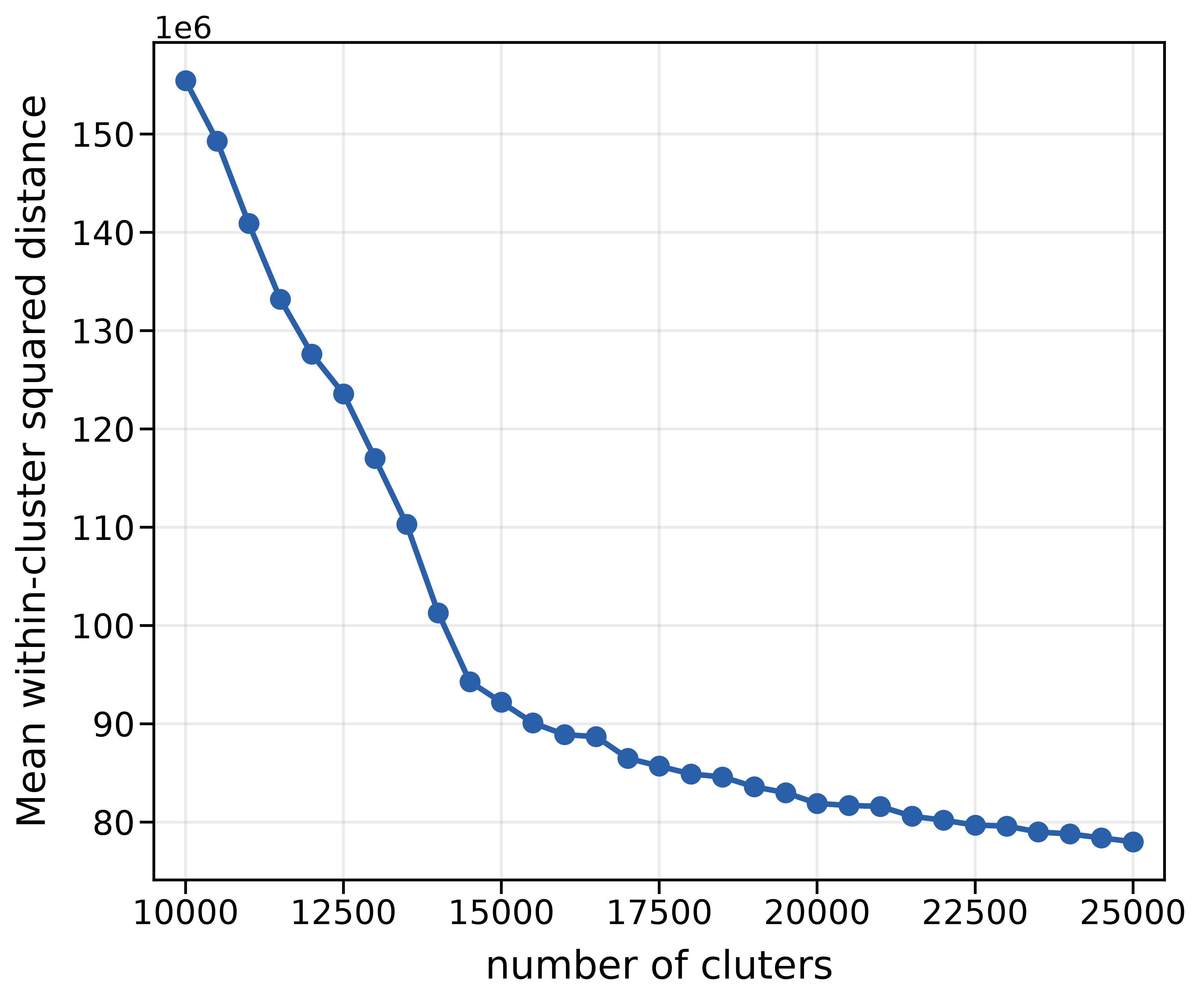}}
    \subcaptionbox{The curve of longitudinal modularity versus the number of clusters.\label{fig:long-mod-k}}{\includegraphics[width=0.45\linewidth]{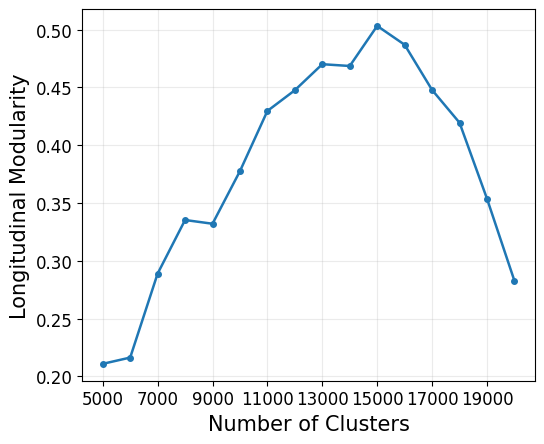}}
    \caption{Selection of the number of clusters on the filtered OpenAlex core graph. The curve in (a) reports within cluster sum of squares (WCSS) under different candidate number of clusters, and (b) reports the longitudinal modularity versus the number of clusters.}
\end{figure}

Since ground-truth dynamic community labels are unavailable for this real-world network, AMI and ARI cannot be directly applied. Instead, we use longitudinal modularity~\cite{brabant2025longitudinal} and within-cluster-sum-of-squares (WCSS)~\cite{thorndike1953belongs} as an external unsupervised quality function for selecting the number of clusters. Longitudinal modularity extends modularity from static graphs to fine-grained temporal interactions and measures whether temporal links occur more frequently within the inferred dynamic communities than expected under a link-stream null model. Its value depends on the temporal network structure, network density, temporal activity patterns, and the assumptions of the chosen null model. Therefore, we use it only for comparing different clustering results on the same OpenAlex CS graph. Specifically, we perform a grid search over candidate values of $K$ and compute the longitudinal modularity and WCSS of the resulting dynamic clustering. As shown in Fig.~\ref{fig:WCSS}, WCSS decreases rapidly as $K$ increases up to approximately 15,000, after which the curve begins to flatten. Besides, Fig.~\ref{fig:long-mod-k} corroborates the point: longitudinal modularity first increases with $K$ and reaches its maximum at $K=15,000$, after which the score generally declines. We therefore use $K=15,000$ for the subsequent community evolution analysis. We also repeated the experiment five times and the average similarity between runs measured by AMI is $0.82$, which suggests the result is replicable.  
\begin{figure}
    \centering
    \includegraphics[width=\linewidth]{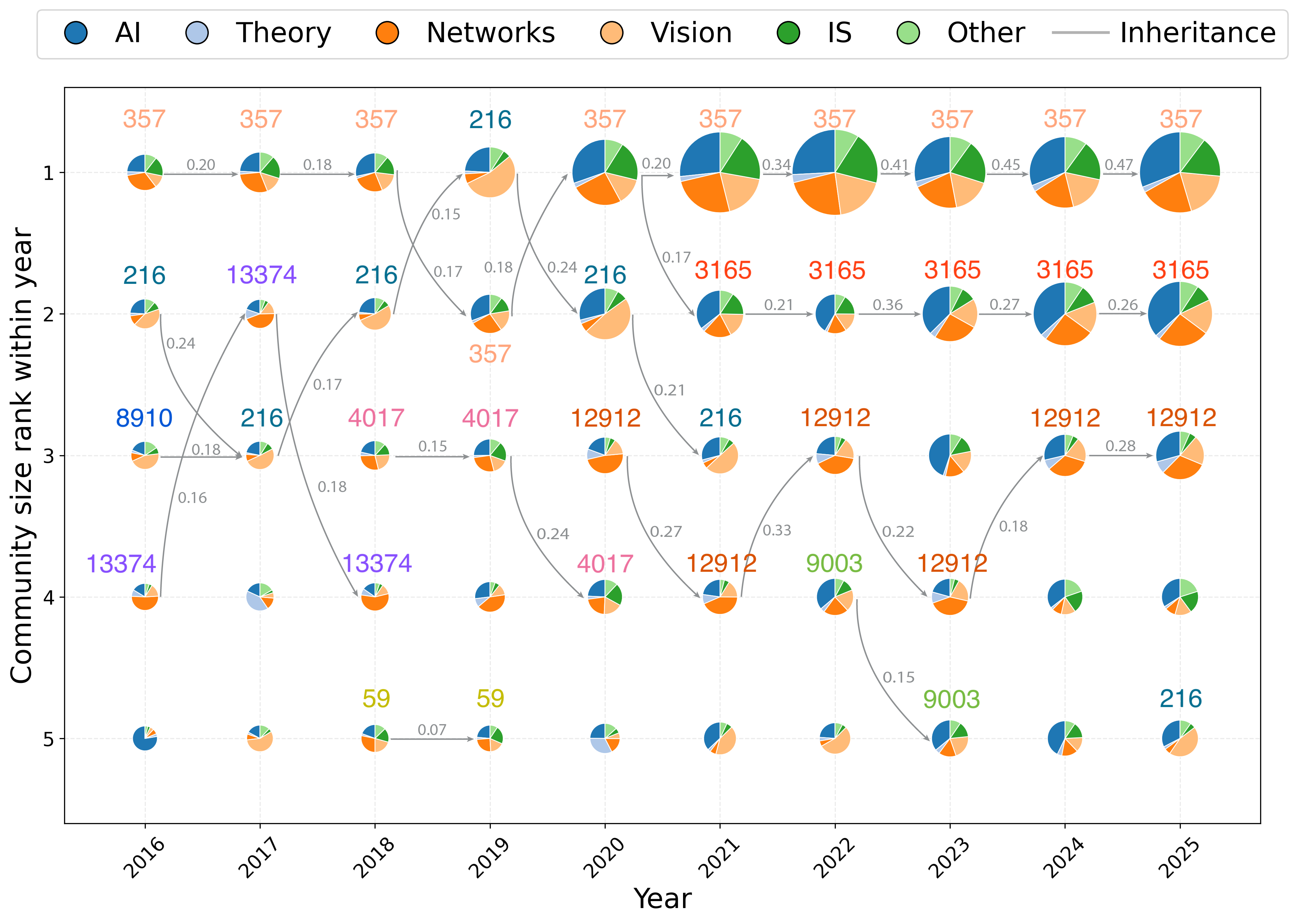}
    \caption{Temporal evolution and inheritance of the top communities in the computer science collaboration network. Each column corresponds to one year from 2016 to 2025, and each row represents the rank of a community by size within that year. The size of each pie chart indicates the relative community size, while the pie slices show the composition of major computer science subfields, including artificial intelligence (AI), Computer Networks and Systems (Networks), Computer Vision and Pattern Recognition (Vision), Information Systems (IS), Signal Processing (SP), and Other. Gray arrows indicate inheritance relationships between communities in adjacent years, with the numbers denoting the corresponding similarity scores measured by Jaccard index.}
    \label{fig:community-inherit}
\end{figure}

To analyse how detected communities evolve over time, we compare communities in consecutive years according to their author overlap. For each community, we identify the community in the following year that shares the largest proportion of authors with it. A larger overlap indicates a stronger temporal inheritance relationship, suggesting that the later community preserves more members from the earlier one. This overlap-based strategy follows the common practice of tracking evolving communities by matching groups across consecutive time steps according to shared membership~\cite{palla2007quantifying}. In Fig.~\ref{fig:community-inherit}, we draw an inheritance link between two communities in adjacent years when their Jaccard similarity is sufficiently large, and annotate the link with the corresponding similarity value. Fig.~\ref{fig:community-inherit} shows that the temporal evolution of the detected communities reveals a dominant and persistent community throughout the observation period. In the early period, a clear merging event can be observed, where two previously distinct community lineages converge into a single successor community. More interestingly, this merged community subsequently splits into two major branches around 2020. Rather than diverging into communities with substantially different disciplinary profiles, the two resulting branches develop into communities with broadly similar subfield compositions, while both expanding considerably in size. These two enlarged communities then persist as the two largest communities in the network throughout the later years. This suggests that the split does not simply reflect disciplinary differentiation. A previously integrated research community appears to separate into two structurally distinct but compositionally similar groups that subsequently grow in parallel. The result highlights how substantial changes in community membership and organisation can occur without correspondingly large changes in their disciplinary composition.
\section{Conclusions}
The proposed framework shows that dynamic clustering can be formulated as what we call ``time-node'' clustering. This is finer-grained than static clustering or clustering based on time windows, with nodes being assigned a community label at any time where they have an interaction with another node. The method was tested on synthetic datasets with a known ground truth against state of the art static and dynamic methods. Our method was better able to predict the community labels than other methods. As might be expected, static methods performed poorly when the ground truth was dynamic although they were faster to run. Of the dynamic methods tested, GenLouvain was competitive when the number of time slices was small and the network changed slowly and occasionally outperformed our methods. However, it was very poor when the network changed quickly and had a large number of time slices. It should be noted that GenLouvain was given an artificial advantage of working with a system that generated communities that remained static within time slices and was given exact information on the length of these time slices and their start points. These assumptions would not be met in real data. LAGO was more consistent in its performance when the network changed quickly, but was consistently worse than our method at replicating the ground truth and would not scale to larger networks.

Our method was then tested on a real network dataset with no known ground truth. The method consistently produced similar community assignments with different starting seeds. It gave useful insights into the underlying dataset, including showing the growth of the largest community over time and the merging of two large communities to become the second largest community. We believe that our method is unique in its capability to analyse how communities change in time without making assumptions about time windows, yet also scale to deal with realistic sizes of datasets. 
\section*{Acknowledgement}
We would like to express thanks to Zhiwei Wang for his contribution in discussing anonymous random walk. We acknowledge the assistance of the ITS Research team at Queen Mary University of London for offering the computational platform.
\bibliographystyle{ACM-Reference-Format}
\bibliography{refs}

@inproceedings{da2020inductive,
title={Inductive representation learning on temporal graphs},
author={Da Xu and Chuanwei Ruan and Evren Korpeoglu and Sushant Kumar and Kannan Achan},
booktitle={International Conference on Learning Representations},
year={2020}
}

@inproceedings{emanuele2020temporal,
    title={Temporal Graph Networks for Deep Learning on Dynamic Graphs},
    author={Emanuele Rossi and Ben Chamberlain and Fabrizio Frasca and Davide Eynard and Federico 
    Monti and Michael Bronstein},
    booktitle={ICML 2020 Workshop on Graph Representation Learning},
    year={2020}
}

@article{bazzi2020framework,
  title={A framework for the construction of generative models for mesoscale structure in multilayer networks},
  author={Bazzi, Marya and Jeub, Lucas GS and Arenas, Alex and Howison, Sam D and Porter, Mason A},
  journal={Physical Review Research},
  volume={2},
  number={2},
  pages={023100},
  year={2020},
  publisher={APS}
}

@article{karrer2011stochastic,
  title={Stochastic blockmodels and community structure in networks},
  author={Karrer, Brian and Newman, Mark EJ},
  journal={Physical Review E—Statistical, Nonlinear, and Soft Matter Physics},
  volume={83},
  number={1},
  pages={016107},
  year={2011},
  publisher={APS}
}

@INPROCEEDINGS{brabant2025discovering,
  author={Brabant, Victor and Bonifati, Angela and Cazabet, Rémy},
  booktitle={2025 IEEE International Conference on Data Mining}, 
  title={Discovering Communities in Continuous-Time Temporal Networks by Optimizing L-Modularity}, 
  year={2025},
  pages={1065-1074}
}

@article{brabant2025longitudinal,
   author = {Victor Brabant and Yasaman Asgari and Pierre Borgnat and Angela Bonifati and Rémy Cazabet},
   issn = {2193-1127},
   issue = {1},
   journal = {EPJ Data Science 2025 14:1},
   month = {2},
   pages = {12-},
   publisher = {SpringerOpen},
   title = {Longitudinal modularity, a modularity for link streams},
   volume = {14},
   year = {2025}
}

@article{newman2006modularity,
   author = {M. E.J. Newman},
   issn = {00278424},
   issue = {23},
   journal = {Proceedings of the National Academy of Sciences of the United States of America},
   month = {6},
   pages = {8577-8582},
   publisher = {National Academy of Sciences},
   title = {Modularity and community structure in networks},
   volume = {103},
   year = {2006}
}

@inproceedings{aynaud2010static,
  title={Static community detection algorithms for evolving networks},
  author={Aynaud, Thomas and Guillaume, Jean-Loup},
  booktitle={8th international symposium on modeling and optimization in mobile, ad hoc, and wireless networks},
  pages={513--519},
  year={2010},
  organization={IEEE}
}

@article{zarayeneh2021delta,
   author = {Neda Zarayeneh and Ananth Kalyanaraman},
   issn = {23274697},
   issue = {2},
   journal = {IEEE Transactions on Network Science and Engineering},
   month = {4},
   pages = {1614-1629},
   publisher = {IEEE Computer Society},
   title = {Delta-Screening: A Fast and Efficient Technique to Update Communities in Dynamic Graphs},
   volume = {8},
   year = {2021}
}

@inproceedings{chong2013incremental,
  title={An incremental batch technique for community detection},
  author={Chong, Wen Haw and Teow, Loo Nin},
  booktitle={Proceedings of the 16th international conference on information fusion},
  pages={750--757},
  year={2013},
  organization={IEEE}
}

@inproceedings{meng2016novel,
  title={A novel dynamic community detection algorithm based on modularity optimization},
  author={Meng, Xiangfeng and Tong, Yunhai and Liu, Xinhai and Zhao, Shuai and Yang, Xianglin and Tan, Shaohua},
  booktitle={2016 7th IEEE international conference on software engineering and service science},
  pages={72--75},
  year={2016},
  organization={IEEE}
}

@article{cordeiro2016dynamic,
   author = {Mário Cordeiro and Rui Portocarrero Sarmento and João Gama},
   issn = {18695469},
   issue = {1},
   journal = {Social Network Analysis and Mining},
   month = {12},
   publisher = {Springer-Verlag Wien},
   title = {Dynamic community detection in evolving networks using locality modularity optimization},
   volume = {6},
   year = {2016}
}

@article{sahu2024df,
  title={DF Louvain: Fast Incrementally Expanding Approach for Community Detection on Dynamic Graphs},
  author={Sahu, Subhajit},
  journal={arXiv preprint arXiv:2404.19634},
  year={2024}
}

@article{mucha2010community,
  title={Community structure in time-dependent, multiscale, and multiplex networks},
  author={Mucha, Peter J and Richardson, Thomas and Macon, Kevin and Porter, Mason A and Onnela, Jukka-Pekka},
  journal={science},
  volume={328},
  number={5980},
  pages={876--878},
  year={2010},
  publisher={American Association for the Advancement of Science}
}

@article{bazzi2016community,
  title={Community detection in temporal multilayer networks, with an application to correlation networks},
  author={Bazzi, Marya and Porter, Mason A and Williams, Stacy and McDonald, Mark and Fenn, Daniel J and Howison, Sam D},
  journal={Multiscale Modeling \& Simulation},
  volume={14},
  number={1},
  pages={1--41},
  year={2016},
  publisher={SIAM}
}

@article{pamfil2019relating,
  title={Relating modularity maximization and stochastic block models in multilayer networks},
  author={Pamfil, A Roxana and Howison, Sam D and Lambiotte, Renaud and Porter, Mason A},
  journal={SIAM Journal on Mathematics of Data Science},
  volume={1},
  number={4},
  pages={667--698},
  year={2019},
  publisher={SIAM}
}

@article{Liu2025DeepDatasets,
    title = {{Deep Temporal Graph Clustering: A Comprehensive Benchmark and Datasets}},
    year = {2025},
    journal = {IEEE Transactions on Pattern Analysis and Machine Intelligence},
    author = {Liu, Meng and Liang, Ke and Wang, Siwei and Hu, Xingchen and Zhou, Sihang and Liu, Xinwang},
    number = {12},
    pages = {11561--11578},
    volume = {47},
    publisher = {IEEE Computer Society},
    doi = {10.1109/TPAMI.2025.3596609},
    issn = {19393539},
    pmid = {40773389}
}

@article{holland1983stochastic,
  title={Stochastic blockmodels: First steps},
  author={Holland, Paul W and Laskey, Kathryn Blackmond and Leinhardt, Samuel},
  journal={Social networks},
  volume={5},
  number={2},
  pages={109--137},
  year={1983},
  publisher={Elsevier}
}

@article{xu2014dynamic,
  title={Dynamic stochastic blockmodels for time-evolving social networks},
  author={Xu, Kevin S and Hero, Alfred O},
  journal={IEEE Journal of Selected Topics in Signal Processing},
  volume={8},
  number={4},
  pages={552--562},
  year={2014},
  publisher={IEEE}
}

@inproceedings{pons2005computing,
  title={Computing communities in large networks using random walks},
  author={Pons, Pascal and Latapy, Matthieu},
  booktitle={International symposium on computer and information sciences},
  pages={284--293},
  year={2005},
  organization={Springer}
}

@article{smiljanic2026community,
  title={Community detection with the map equation and infomap: Theory and applications},
  author={Smiljani{\'c}, Jelena and Bl{\"o}cker, Christopher and Holmgren, Anton and Edler, Daniel and Neuman, Magnus and Rosvall, Martin},
  journal={ACM Computing Surveys},
  volume={58},
  number={7},
  pages={1--34},
  year={2026},
  publisher={ACM New York, NY}
}

@article{von2007tutorial,
  title={A tutorial on spectral clustering},
  author={Von Luxburg, Ulrike},
  journal={Statistics and computing},
  volume={17},
  number={4},
  pages={395--416},
  year={2007},
  publisher={Springer}
}

@article{aslak2018constrained,
  title={Constrained information flows in temporal networks reveal intermittent communities},
  author={Aslak, Ulf and Rosvall, Martin and Lehmann, Sune},
  journal={Physical Review E},
  volume={97},
  number={6},
  pages={062312},
  year={2018},
  publisher={APS}
}

@article{Pareja2019EvolveGCN,
    title = {{EvolveGCN: Evolving Graph Convolutional Networks for Dynamic Graphs}},
    year = {2019},
    author = {Pareja, Aldo and Domeniconi, Giacomo and Chen, Jie and Ma, Tengfei and Suzumura, Toyotaro and Kanezashi, Hiroki and Kaler, Tim and Schardl, Tao B. and Leiserson, Charles E.},
    month = {11},
    arxivId = {1902.10191}
}

@article{sankar2018dynamic,
  title={Dynamic graph representation learning via self-attention networks},
  author={Sankar, Aravind and Wu, Yanhong and Gou, Liang and Zhang, Wei and Yang, Hao},
  journal={arXiv preprint arXiv:1812.09430},
  year={2018}
}

@inproceedings{nguyen2018continuous,
    address = {Lyon, France},
    title = {Continuous-{Time} {Dynamic} {Network} {Embeddings}},
    doi = {10.1145/3184558.3191526},
    language = {en},
    urldate = {2025-01-24},
    booktitle = {Companion of the {The} {Web} {Conference} 2018 on {The} {Web} {Conference} 2018 - {WWW} '18},
    publisher = {ACM Press},
    author = {Nguyen, Giang Hoang and Lee, John Boaz and Rossi, Ryan A. and Ahmed, Nesreen K. and Koh, Eunyee and Kim, Sungchul},
    year = {2018},
    pages = {969--976},
}

@article{makarov2021temporal,
  title={Temporal graph network embedding with causal anonymous walks representations},
  author={Makarov, Ilya and Savchenko, Andrey and Korovko, Arseny and Sherstyuk, Leonid and Severin, Nikita and Mikheev, Aleksandr and Babaev, Dmitrii},
  journal={arXiv preprint arXiv:2108.08754},
  year={2021}
}

@inproceedings{zuo2018embedding,
  title={Embedding temporal network via neighborhood formation},
  author={Zuo, Yuan and Liu, Guannan and Lin, Hao and Guo, Jia and Hu, Xiaoqian and Wu, Junjie},
  booktitle={Proceedings of the 24th ACM SIGKDD international conference on knowledge discovery \& data mining},
  pages={2857--2866},
  year={2018}
}

@inproceedings{liumeng2021inductive,
  title={Inductive representation learning in temporal networks via mining neighborhood and community influences},
  author={Liu, Meng and Liu, Yong},
  booktitle={Proceedings of the 44th International ACM SIGIR Conference on Research and Development in Information Retrieval},
  pages={2202--2206},
  year={2021}
}

@article{priem2022openalex,
  title={OpenAlex: A fully-open index of scholarly works, authors, venues, institutions, and concepts},
  author={Priem, Jason and Piwowar, Heather and Orr, Richard},
  journal={arXiv preprint arXiv:2205.01833},
  year={2022}
}

@article{palla2007quantifying,
  title={Quantifying social group evolution},
  author={Palla, Gergely and Barab{\'a}si, Albert-L{\'a}szl{\'o} and Vicsek, Tam{\'a}s},
  journal={Nature},
  volume={446},
  number={7136},
  pages={664--667},
  year={2007},
  doi={10.1038/nature05670}
}

@manual{king2017apocrita,
  title= {Apocrita - High Performance Computing Cluster for Queen Mary University of London},
  author= {King, Thomas and Butcher, Simon and Zalewski, Lukasz},
  month= mar,
  year= 2017,
  doi= {10.5281/zenodo.438045}
}

@article{blondel2008fast,
  title={Fast unfolding of communities in large networks},
  author={Blondel, Vincent D and Guillaume, Jean-Loup and Lambiotte, Renaud and Lefebvre, Etienne},
  journal={Journal of statistical mechanics: theory and experiment},
  volume={2008},
  number={10},
  pages={P10008},
  year={2008}
}

@article{cai2018comprehensive,
  title={A comprehensive survey of graph embedding: Problems, techniques, and applications},
  author={Cai, Hongyun and Zheng, Vincent W and Chang, Kevin Chen-Chuan},
  journal={IEEE transactions on knowledge and data engineering},
  volume={30},
  number={9},
  pages={1616--1637},
  year={2018},
  publisher={IEEE}
}

@article{ng2001spectral,
  title={On spectral clustering: Analysis and an algorithm},
  author={Ng, Andrew and Jordan, Michael and Weiss, Yair},
  journal={Advances in neural information processing systems},
  volume={14},
  year={2001}
}

@inproceedings{vinh2009information,
  title={Information theoretic measures for clusterings comparison: is a correction for chance necessary?},
  author={Vinh, Nguyen Xuan and Epps, Julien and Bailey, James},
  booktitle={Proceedings of the 26th annual international conference on machine learning},
  pages={1073--1080},
  year={2009}
}

@article{danon2005comparing,
  title={Comparing community structure identification},
  author={Danon, Leon and Diaz-Guilera, Albert and Duch, Jordi and Arenas, Alex},
  journal={Journal of statistical mechanics: Theory and experiment},
  volume={2005},
  number={09},
  pages={P09008--P09008},
  year={2005}
}

@article{rand1971objective,
  title={Objective criteria for the evaluation of clustering methods},
  author={Rand, William M},
  journal={Journal of the American Statistical association},
  volume={66},
  number={336},
  pages={846--850},
  year={1971},
  publisher={Taylor \& Francis}
}

@article{hubert1985comparing,
  title={Comparing partitions journal of classification 2 193--218},
  author={Hubert, L and Arabie, P},
  journal={Google Scholar},
  volume={193},
  year={1985}
}

@inproceedings{liu2024revisiting,
  title={Revisiting modularity maximization for graph clustering: A contrastive learning perspective},
  author={Liu, Yunfei and Li, Jintang and Chen, Yuehe and Wu, Ruofan and Wang, Ericbk and Zhou, Jing and Tian, Sheng and Shen, Shuheng and Fu, Xing and Meng, Changhua and others},
  booktitle={Proceedings of the 30th ACM SIGKDD Conference on Knowledge Discovery and Data Mining},
  pages={1968--1979},
  year={2024}
}

@article{liu2026survey,
  title={A survey of deep graph clustering: Taxonomy, challenge, application, and open resource},
  author={Liu, Yue and Xia, Jun and Wu, Benyu and Zhou, Sihang and Yang, Xihong and Liang, Ke and Fan, Chenchen and Zhuang, Yan and Yu, Guoxian and Li, Stan Z and others},
  journal={IEEE Transactions on Knowledge and Data Engineering},
  year={2026},
  publisher={IEEE}
}

@inproceedings{wang2017mgae,
  title={Mgae: Marginalized graph autoencoder for graph clustering},
  author={Wang, Chun and Pan, Shirui and Long, Guodong and Zhu, Xingquan and Jiang, Jing},
  booktitle={Proceedings of the 2017 ACM on Conference on Information and Knowledge Management},
  pages={889--898},
  year={2017}
}

@article{wang2019attributed,
  title={Attributed graph clustering: A deep attentional embedding approach},
  author={Wang, Chun and Pan, Shirui and Hu, Ruiqi and Long, Guodong and Jiang, Jing and Zhang, Chengqi},
  journal={arXiv preprint arXiv:1906.06532},
  year={2019}
}

@inproceedings{bo2020structural,
  title={Structural deep clustering network},
  author={Bo, Deyu and Wang, Xiao and Shi, Chuan and Zhu, Meiqi and Lu, Emiao and Cui, Peng},
  booktitle={Proceedings of the web conference 2020},
  pages={1400--1410},
  year={2020}
}

@inproceedings{hassani2020contrastive,
  title={Contrastive multi-view representation learning on graphs},
  author={Hassani, Kaveh and Khasahmadi, Amir Hosein},
  booktitle={International conference on machine learning},
  pages={4116--4126},
  year={2020},
  organization={PMLR}
}

@inproceedings{liu2022deep,
  title={Deep graph clustering via dual correlation reduction},
  author={Liu, Yue and Tu, Wenxuan and Zhou, Sihang and Liu, Xinwang and Song, Linxuan and Yang, Xihong and Zhu, En},
  booktitle={Proceedings of the AAAI conference on artificial intelligence},
  volume={36},
  number={7},
  pages={7603--7611},
  year={2022}
}

@inproceedings{yang2023cluster,
  title={Cluster-guided contrastive graph clustering network},
  author={Yang, Xihong and Liu, Yue and Zhou, Sihang and Wang, Siwei and Tu, Wenxuan and Zheng, Qun and Liu, Xinwang and Fang, Liming and Zhu, En},
  booktitle={Proceedings of the AAAI conference on artificial intelligence},
  volume={37},
  number={9},
  pages={10834--10842},
  year={2023}
}

@article{yuan2025temporal,
  title={Temporal community detection and analysis with network embeddings},
  author={Yuan, Limengzi and Zhang, Xuanming and Ke, Yuxian and Lu, Zhexuan and Li, Xiaoming and Liu, Changzheng},
  journal={Mathematics},
  volume={13},
  number={5},
  pages={698},
  year={2025},
  publisher={MDPI}
}

@inproceedings{agdur2025approximating,
  title={Approximating temporal modularity on graphs of small underlying treewidth},
  author={Agdur, Vilhelm and Enright, Jessica and Larios-Jones, Laura and Meeks, Kitty and Skerman, Fiona and Yates, Ella},
  booktitle={International Symposium on Algorithmics of Wireless Networks},
  pages={17--31},
  year={2025},
  organization={Springer}
}

@article{abboud2020surprising,
  title={The surprising power of graph neural networks with random node initialization},
  author={Abboud, Ralph and Ceylan, Ismail Ilkan and Grohe, Martin and Lukasiewicz, Thomas},
  journal={arXiv preprint arXiv:2010.01179},
  year={2020}
}

@article{duong2019node,
  title={On node features for graph neural networks},
  author={Duong, Chi Thang and Hoang, Thanh Dat and Dang, Ha The Hien and Nguyen, Quoc Viet Hung and Aberer, Karl},
  journal={arXiv preprint arXiv:1911.08795},
  year={2019}
}

@inproceedings{charikar2002finding,
  title={Finding frequent items in data streams},
  author={Charikar, Moses and Chen, Kevin and Farach-Colton, Martin},
  booktitle={International colloquium on automata, languages, and programming},
  pages={693--703},
  year={2002},
  organization={Springer}
}

@inproceedings{chen2020simple,
  title={A simple framework for contrastive learning of visual representations},
  author={Chen, Ting and Kornblith, Simon and Norouzi, Mohammad and Hinton, Geoffrey},
  booktitle={International conference on machine learning},
  pages={1597--1607},
  year={2020},
  organization={PmLR}
}

@inproceedings{wang2021understanding,
  title={Understanding the behaviour of contrastive loss},
  author={Wang, Feng and Liu, Huaping},
  booktitle={Proceedings of the IEEE/CVF conference on computer vision and pattern recognition},
  pages={2495--2504},
  year={2021}
}

@article{thorndike1953belongs,
  title={Who belongs in the family?},
  author={Thorndike, Robert L},
  journal={Psychometrika},
  volume={18},
  number={4},
  pages={267--276},
  year={1953},
  publisher={Cambridge University Press \& Assessment}
}




\end{document}